\RequirePackage{fix-cm}
\documentclass[twocolumn]{svjour3}          
\smartqed  
\usepackage{graphicx, amsmath, amssymb}
\usepackage{subfigure}
\usepackage{cite}
\newcommand{\intd}{\,\mbox{d}}
\newcommand{\mb}[1]{\mbox{\boldmath $#1$}}
\newcommand{\pd}[2]{\frac{\partial #1}{\partial #2}}

\renewcommand{\tens}[1]{\boldsymbol{#1}} 
\newcommand{\abs}[1]{\left\|#1\right\|}

\newcommand{\inv}[1]{\frac{1}{#1}}

\newcommand{\hlm}{\hat{\tens\lambda}}
\newcommand{\hu}{\hat{\mb u}}
\newcommand{\dhu}{\dot{\hat{\mb u}}}
\newcommand{\ddhu}{\ddot{\hat{\mb u}}}
\newcommand{\hq}{\hat{\mb q}}
\newcommand{\dhq}{\dot{\hat{\mb q}}}
\newcommand{\ddhq}{\ddot{\hat{\mb q}}}

\newcommand{\intV}[2]{\int_{#1}#2\intd V}

\begin{document}

\title{Computing Arlequin coupling coefficient for concurrent FE-MD approaches\thanks{We would like to thank the German Science Foundation (DFG) for the financial support, through the project DFG SH 581/2-1.}
}
\subtitle{}


\author{Wenzhe Shan        \and
        Udo Nackenhorst 
}


\institute{W. Shan \at
              Institute for Advanced Study, Nanchang University, Jiangxi, China \\
              Tel.: +86-791-83968935\\              
              \email{shan@ncu.edu.cn}           
}

\date{Received: date / Accepted: date}

\maketitle

\begin{abstract}
Arlequin coupling coefficient is essential for concurrent FE-MD models with overlapping domains, but the calculation of its value is quite difficult when the geometry of the coupling region is complicated. In this work, we introduce a general procedure for the preprocessing of a concurrent FE-MD model, given that the mesh and atoms have already been created. The procedure is independent of the geometry of the coupling region and can be used for both 2D and 3D problems. The procedure includes steps of determining the relative positions of atoms inside the FE elements in the coupling region, as well as computing the Arlequin coupling coefficient for an arbitrary point inside the coupling region or on its boundary. Two approaches are provided for determining the coefficient: the direct approach and the temperature approach.
\end{abstract}

\section{Introduction}
\label{intro}
Concurrent FE-MD coupling models have found their applications for mesoscale problems which contain special regions where explicit atomistic model is desired \cite{fish2006, fish2007}, while the size of the entire region is too large to be fully modeled by atoms. Typical approaches include: the bridging domain method \cite{xiao2004}, the weak coupling method \cite{dhia1998, dhia2008, bauman2008, fackeldey2009}and the Quasicontinuum (QC) method \cite{tadmor1996a}. In order to keep the energy and momentum consistency in the coupling region, interpolation of energy is required for the overlapping part of the atomistic and continuum models. A commonly used interpolation approach is the Arlequin method \cite{xiao2004, dhia1998, dhia2004, dhia2005, dhia2008}, where the energy in the coupling region takes the form of
\begin{equation}
 \Pi = \alpha\Pi^{FE} + (1-\alpha)\Pi^{MD},
 \label{eq:def_coupling_energy}
\end{equation}
Despite its simple concept, the implementation for computing the coupling coefficient $\alpha$ by the Arlequin method for general 2D and 3D problems are quite involved. To the best of the author's knowledge, no systematic approach has been presented in literature yet. In this contribution, we would like to introduce a relatively simple but robust approach for such task.\\

This contribution is structured in the following way: the general form of equations of motion for the concurrent FE-MD approaches is first reviewed, so that the difficulty of the implementation can be clearly identified. The main challenge of obtaining the Arlequin coupling coefficient lies in two-fold: determining the relative position of the atoms inside the FE elements and calculating the coefficient at desired positions, e.g. the Gaussian points of FE elements and atoms in the coupling region. The procedure for both tasks will detailed in two sections afterwards. Two approaches will be introduced for computing the coupling coefficient: the direct approach and the temperature approach. At the end, the influence of the coupling coefficient on the dynamic behavior of a FE-MD model is investigated by simulating the propagation of a Gaussian wave in a 2D linear FE-MD model. The behavior of the model with the coupling coefficient computed from different approaches are compared with each other, as well as to the cases when no scaling and constant coupling coefficient is used.

\section{Equations of Motions}
\label{sec:EOM}
Since our focus in this work is not the material modeling, without loss of generality, we will use linear models as the template for the equations of motion. Moreover, in the following discussion, we only consider the formulation in the coupling region. \\

The mechanical energy of the coupling region can be written as
\begin{equation}
  \begin{aligned}
  \Pi & = \mathcal K + \mathcal U\\
  \mathcal K & = \mathcal K^{FE} + \mathcal K^{MD}\\
  \mathcal U & = \mathcal U^{FE} + \mathcal U^{MD}
  \end{aligned}
  \label{eq:def_energies}
\end{equation}
where $\mathcal K$ denotes the kinetic energy and $\mathcal U$ denotes the potential energy. The kinetic energy of the FE model and the MD model can be written as
\begin{equation}
 \begin{aligned}
    \mathcal K^{FE} & = \inv{2}\dhu^T\left(\intV{\Omega_{cp}}{\alpha(\mb X)\mb N_u^T\mb N_u}\right)\dhu = \inv{2}\dhu^T\tilde{\mb M}_u\dhu\\
  \mathcal K^{MD} & = \inv{2}\sum_i\dot{\mb q}_i^T\left(1 -\alpha_i\right)m_i\dot{\mb q}_i = \inv{2}\dhq^T\tilde{\mb M}_q\dhq
 \end{aligned}
\label{eq:def_energies_K}
\end{equation}
where $\dhu$ is the global vector that contains all nodal velocities of the FE model and $\dhq$ contains the velocities of atoms of the MD model. $\mb N_u$ is the shape function matrix of FE model and $m_i$ is the mass of atom $i$. The mass matrices $\tilde{\mb M}_u$ and $\tilde{\mb M}_q$ are scaled from the original ones by the coupling coefficient $\alpha$, which has the boundary values:
\begin{equation}
  \alpha(\mb x) = \left\{\begin{array}{ll}
                          0, &\hbox{on the MD side}\\
                          1, &\hbox{on the FE side}
                         \end{array}\right.
 \label{eq:alpha_limit}
\end{equation}
Similarly, the potential energies, for linear models, can be written in quadratic forms with the scaled stiffness matrices $\tilde{\mb K}_u$ and $\tilde{\mb K}_q$, as
\begin{equation}
 \begin{aligned}
  \mathcal U^{FE} & = \inv{2}\hu^T\left(\intV{\Omega_{cp}}{\alpha(\mb x)\mb B^T\mathbb{C}\mb B}\right)\hu = \inv{2}\hu^T\tilde{\mb K}_u\hu\\
  \mathcal U^{MD} & = \sum_i\sum_j\mb q_i^T\frac{\left(2 -\alpha_i - \alpha_j\right)}{2}\mb K^{MD}_{ij}\mb q_j = \inv{2}\hq^T\tilde{\mb K}_q\hq
 \end{aligned}
  \label{eq:def_energies_U}
\end{equation}
where $\mb B$ is the strain-displacement matrix \cite{wriggers2008}, $\mathbb{C}$ is the consistent elastic tensor \cite{shan2009d} and $\mb K^{MD}_{ij}$ represents the 2nd order derivative of the potential between atom $i$ and $j$ at the equilibrium configuration \cite{shan2009d}, assuming pair potential is used. Vector $\hu$ contains the nodal displacements of the FE model and $\hq$ the atomic displacements of the MD model. \\

For the weak-coupling method (WCM), the atomic displacements of the MD model are first interpolated by a continuous function and then mapped to the FE model adopting L2 projection  \cite{fackeldey2009}. We have
\begin{equation}
\begin{aligned}
 &g(\mb u, \mb q, \tens\lambda) = \intV{\Omega_{cp}}{\tens\lambda^T\left(\mb u(\mb X) - \mb q(\mb x)\right)} = \mb 0\\
 &\Longrightarrow \left(\intV{\Omega_{cp}}{\mb N_u^T\mb N_u}\right)\hu   = \left(\intV{\Omega_{cp}}{\mb N_u^T\tens\phi_q}\right)\hq
\end{aligned}
\label{eq:def_WC}
\end{equation}
where the $\tens\phi_q$ is the shape function matrix that interpolates the discrete atomistic displacements $\hq$ into a continuous representation $\mb q(\mb x)$ and we in addition assume that the Lagrangian multiplier $\tens\lambda$ is descretized by the same shape function matrix as the FE displacement $\mb u(\mb x)$. In such case, the MD solution is decomposed into two orthogonal parts: its projection to the FE model and an error part\cite{shan2013}. For the bridging domain method (BDM), the MD displacements are considered to be the interpolation of FE nodal displacements\cite{xiao2004}. The Lagrangian multiplier term and the displacement constraint can be written as
\begin{equation}
	\begin{aligned}
		&g(\mb u, \mb q, \tens\lambda) = \hlm^T\left(\hq - \mathbb N\hu\right) = 0\\
		&\Longrightarrow \hq = \mathbb N\hu
	\end{aligned}
	\label{eq:def_BDM}
\end{equation}
where $\mathbb N$ is a matrix with components of the FE shape functions evaluated at atomic positions \cite{shan2013}.\\

By observing the above results, the displacement constraints can be written in a common form as
\begin{equation}
	\mb W_u\hu = \mb W_q\hq,
\end{equation}
with
\begin{equation}
	\begin{aligned}
	\mb W_u & = \left\{\begin{array}{lr}
			\intV{\Omega_{cp}}{\mb N_u^T\mb N_u}&\hbox{for WCM}\\
			\mathbb N&\hbox{for BDM}
	\end{array}\right.\\
	\mb W_q &= \left\{\begin{array}{lr}
			\intV{\Omega_{cp}}{\mb N_u^T\tens\phi_q} & \hbox{for WCM}\\
			\mb I & \hbox{for BDM}\end{array}\right.
	\end{aligned}
	\label{eq:def_Wu_Wq}
\end{equation}
where $\mb I$ denotes the identity matrix. As both matrices $\mb W_u$ and $\mb W_q$ are time-independent, the velocity and acceleration obeys the same constraint,
\begin{equation}
	\mb W_u\dhu = \mb W_q\dhq\;\;\hbox{and}\;\;	\mb W_u\ddhu = \mb W_q\ddhq.
\end{equation}

With the above definitions, the equations of motion for the coupling region can be written as
\begin{equation}
	\begin{aligned}
		\tilde{\mb M}_u\ddhu & = -\tilde{\mb K}_u\hu - \mb W_u^T\hlm\\
		\tilde{\mb M}_q\ddhq & = -\tilde{\mb K}_q\hq + \mb W_q^T\hlm\\
		\mb W_u\dhu & = \mb W_q\dhq.
	\end{aligned}
	\label{eq:eom_general}
\end{equation}
For details of the derivation, it is referred to \cite{shan2013}. Theoretically, it is equivalent to apply the constraint on $\hu$, $\dhu$ or $\ddhu$, but due to numerical reasons, the constraint is usually applied to velocities \cite{xiao2004}. \\

After defining the equations of motion, the implementation difficulties can now be clearly identified. In Eqn.\eqref{eq:def_Wu_Wq}, for the BDM, the coupling matrix $\mb W_u$ is the shape function matrix of the FE model evaluated at the positions of atoms in the coupling region. On the other hand, for the WCM, the coupling matrix $\mb W_q$ is the integration of the product of the FE shape function matrix $\mb N_u$ and the MD interpolation function matrix $\tens\phi_q$. $\mb N_u$ is defined in the FE model while $\tens\phi_q$ defined in the MD model. Delaunay triangulation can be used to connect atoms in the coupling region, creating a MD mesh for the integration. $\mb N^T_u\tens\phi_q$ is then evaluated at Gaussian quadrature points in the MD mesh. To evaluate $\mb N_u$ at the MD quadrature points, the iso-parametric coordinates of the MD quadrature points within corresponding FE element must be determined. This can be accomplished by interpolating the iso-parametric coordinates of the nodes, which are the atoms, of the MD mesh within the corresponding FE element. Therefore, we need to find out the relative positions of atoms inside the FE model in the coupling region. Moreover, the evaluation of the mass matrices and stiffness matrices in Eqn.\eqref{eq:eom_general} requires the value of the Arlequin coupling coefficient $\alpha$. Two approaches for calculating $\alpha$ for an arbitrary point in the coupling region will be introduced in Sec. \ref{sec:alpha}.

\section{Atomic Positions in FE Models}
The calculation of the relative positions of atoms within the FE elements within the coupling region is not only essential for evaluating the coupling matrices $\mb W_u$ and $\mb W_q$, but also necessary for identifying the coupling region and for determining the coupling coefficient. The coupling region can be defined as the union of FE elements which contain  atoms. In this contribution, the relative positions of atoms in the FE elements are their iso-parametric coordinates. The detailed procedures for computing the iso-parametric coordinates of atoms inside a given element are detailed in the appendix of this work. Once the iso-parametric coordinates of atoms have been determined, the coupling coefficient at atomic positions can be interpolated from the nodal coordinates of the FE elements. \\

The procedure for computing the iso-parametric coordinates of a point in an element is called the \emph{inverse iso-parametric mapping} in this work. Whether this point is inside or outside the element can be determined from values of the obtained iso-parametric coordinates. Such status will be called \emph{in/out status}. In a brute-force approach, we need to check the iso-parametric coordinates for all atoms of the MD model with respect to all FE elements. The computational cost of such procedure is linear to the product of the number of atoms and the number of FE elements. If no assumption is made on the geometry of the coupling region, it is the only universal approach to identify FE elements that contain atoms. On the other hand, the inverse iso-parametric mapping is an expensive operation by itself. Therefore, in order to improve the efficiency, it is better to localize the search area so that the inverse iso-parametric mapping of an given atom needs to be carried out with respect to only few elements, instead of the entire FE model.\\

\begin{figure}[htp]
	\centering
	\includegraphics[width = .35\textwidth]{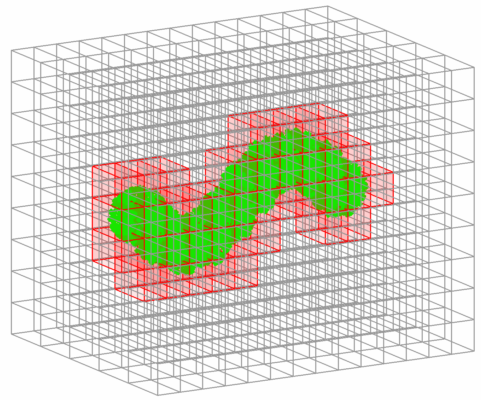}
	\caption{FE elements containing atoms of the MD model, identified as the overlapping domain.}
	\label{fig:eg_overlap_domain}
\end{figure}

The localization approach used by us is similar to the neighbor-list search algorithm \cite{rapaport2004} for MD simulations. It can be summarized into the following steps:
\begin{enumerate}
	\item Divide the simulation region into uniform cells. The minimum size of the cell should be larger than the size of the largest FE element. Such constraint is designed for avoiding having cells with no FE nodes inside.
	\item Calculate the cell coordinates $\mb x_c$ of all atoms and FE nodes by:
		\begin{equation}
			\{x_c\}_i = \left\lfloor \frac{x_i}{\{l_c\}_i^{-1}}\right\rfloor,\hspace{5pt}i\;\in\;\left\{1, 2, 3\right\}
		\end{equation}
		where ${l_c}_i$ is the length of the cell in the i-th direction.
	\item Go through each FE element and perform the inverse iso-parametric mapping only on atoms with the same cell coordinates as any of its nodes. To further improve the efficiency, all atoms found to be within the element shall be marked as "unavailable" for future search.
\end{enumerate}
In the first step described above, a lower bound on the size of the search cell is used to make sure that each cell contains at least one FE node. This is because, during the search, the inverse iso-parametric mapping is performed to each atom with respect to FE elements containing at least one node which shares the same cell coordinates as the atom. Therefore, any cell contains atoms but no FE node would be problematic. One example of detecting the overlapping domain of an irregular shape is shown in Fig.~\ref{fig:eg_overlap_domain}. \\

\section{Arlequin Coupling Coefficient}
\label{sec:alpha}
The Arlequin coupling coefficient interpolates the energies of different models in the overlapping domain. It should be $1$ on one side of the coupling boundary and $0$ on the other side. In this work, it is always denoted by $\alpha$, and, without loss of generality, we consider it reaches identity on the FE side of the coupling boundary and vanishes at the MD side, as defined in Eqn.\eqref{eq:alpha_limit}. \\

From definitions of the kinetic energies, Eqn.\eqref{eq:def_energies_K}, and the definitions of the potential energies, Eqn.\eqref{eq:def_energies_U}, in the coupling region, the evaluation of the Arlequin coupling coefficient is required at each integration point (Gaussian quadrature points in this work) in the FE elements and at each atomic position in the overlapping region. Such evaluation is quite straightforward for 1D case, or for the case where the shapes of the MD region and the coupling region are highly symmetric, such as square, cube, circle and sphere. But once the shape of MD domain becomes irregular, the calculation of the Arlequin coupling coefficient for an arbitrary point becomes quite involved. In this section, we introduce two robust approaches for such situations. \\

Two common steps for both approaches are first: \emph{coupling boundary search} and \emph{boundary ray-tracing}.

\subsection{Coupling Boundary Search}
\label{sec:ix_bc}
The goal of the coupling boundary search is to find the FE nodes on the boundary of the coupling region. We also assume that the boundary can be divided into two distinct sides: the inner side - the side connected to the pure MD region; and the outer side - the side connected to the pure FE region. \\

\begin{figure}[htp]
	\centering
	\subfigure[2D]{\includegraphics[width = .2\textwidth]{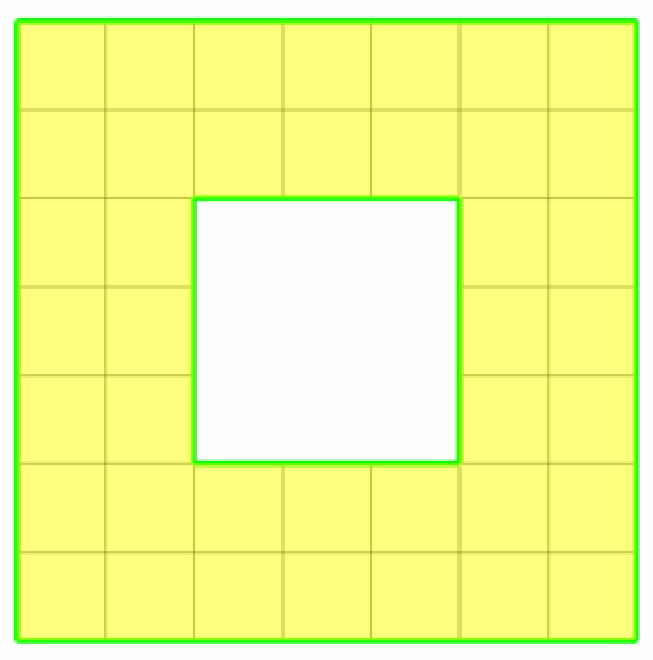}}\hspace{10pt}
	\subfigure[3D]{\includegraphics[width = .2\textwidth]{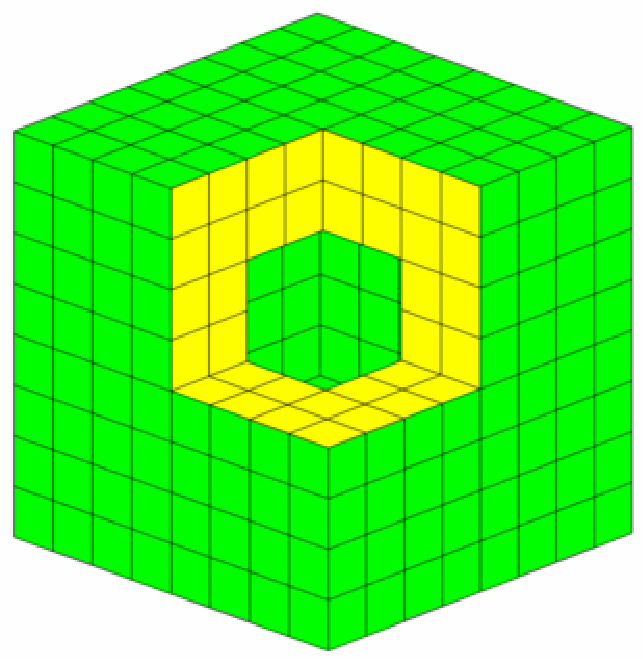}}
	\caption{FE elements in the coupling region (yellow) and the boundary (green). The rest of the model and atoms are not plotted}
	\label{fig:eg_cp_bc}
\end{figure}

Searching for the coupling boundary is equivalent to searching for the outer boundary of the coupling region formed by FE elements containing atoms. The basic idea used in this work is simple: looking for the surface elements on the boundary of the coupling boundary. More specifically, they are:
\begin{itemize}
	\item Nodes, for 1D models;
	\item Edges, for 2D models;
	\item Faces (Triangles or quadrilaterals), for 3D models.
\end{itemize}
When we break each element into a list of objects: for example, a triangle with nodes $[p_1, p_2, p_3]$ into 3 edges: $[p_1, p_2]$, $[p_2, p_3]$ and $[p_3, p_1]$, a tetrahedron $[p_1, p_2, p_3, p_4]$ into four triangles $[p_1, p_2, p_3]$, $[p_1, p_2, p_4]$, $[p_2, p_3, p_4]$ and $[p_3, p_1, p_4]$, a hexahedron element into 6 quadrilateral elements, so on so forth; If we save all the objects into a list after sorting the nodal indices in each object in a fixed order, either ascending or descending. Then \emph{we can determine the number of FE elements an object is shared by, by counting its repetitions in the list}. Since a surface element is only shared by one FE element, all the surface elements will appear only once in the list. They can be identified by searching for the unique entries in the list: the first time taking the indices of the entries which are the first of their repetitions, while the 2nd time taking the indices of the last of their repetitions. The common indices from these two searches are then the indices of the unique entries in the list, as it is only possible for a unique entry to have the index whose first occurrence in the list equals its last. Two examples of the identified boundary of the coupling region are shown in Fig.~\ref{fig:eg_cp_bc}

\subsection{Boundary Ray-Tracing}
\begin{figure}[htp]
	\centering
	\includegraphics[width = .4\textwidth]{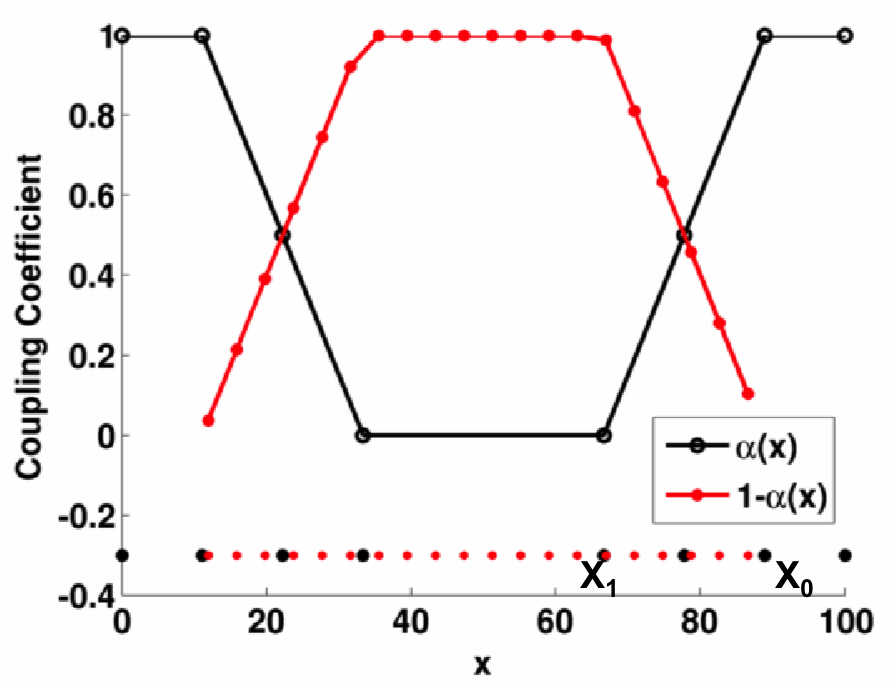}
	\caption{Linear Arlequin coupling coefficients for the FE model (black) and for the MD model (red): $x_0$ is the coupling boundary on the FE side and $x_1$ is that on the MD side.}
	\label{fig:alpha_1D}
\end{figure}

Considering linear Arlequin coupling coefficient  \cite{xiao2004}, for 1D problems it can be computed by:
\begin{equation}
	\alpha(x) = \frac{|x - x_0|}{|x_1 - x_0|},\hspace{10pt} \alpha\,\in\,\left[0, 1\right]
	\label{eq:alpha_1D}
\end{equation}
where $x$ is coordinate of a point in the coupling region, $x_0$ is the coordinate of boundary of the coupling region on the MD side and $x_1$ is the coordinate on the FE side, as shown in Fig.~\ref{fig:alpha_1D}.\\

The coefficient for the 2D and 3D problems can be defined in the same way, as
\begin{equation}
	\alpha(\mb{x}) = \frac{\abs{\mb{x} - \mb{x}_0}}{\abs{\mb{x}_1 - \mb{x}_0}},\hspace{10pt} \alpha\,\in\,\left[0, 1\right].
	\label{eq:alpha_2D_3D}
\end{equation}
However, unlike the 1D case, all the positions used in above equation are vectors instead of scalars. Moreover, the choices of the boundary points $\mb x_0$ and $\mb x_1$ are not unique anymore. So the first thing we need to do is to find a way to define them. From now on, without special notification, $\mb x_0$ and $\mb x_1$ will always denote the boundary points mentioned above. \\

To determine $\mb x_0$ and $\mb x_1$ for a given point $\mb x$ in the coupling region, we need a line that passes through $\mb x$, then its intersection with the inner (MD side) boundary of coupling region is $\mb x_0$ and its intersection with the outer (FE side) boundary is $\mb x_1$. To define this line, we need a second point inside the MD region. It is given the name \textbf{\emph{anchor point} } in this work, denoted by $\mb x_a$\\

Once the anchor point is given, we can define a \emph{ray} starting from the anchor point and pointing in the direction of $\mb x$. All we need to do is to find the intersections of this ray with the boundary of the coupling region. In section \ref{sec:ix_bc}, we introduced the approach of identifying the boundary of the coupling region by identifying surface elements of the FE elements containing atoms. For 2D and 3D problems, we have three types of boundary elements: bar element for 2D models and triangle and quadrilateral elements for 3D problems. Furthermore, we can always divide a quadrilateral element with nodes $[p_1, p_2, p_3, p_4]$ into two triangles $[p_1, p_2, p_3]$ and $[p_1, p_3, p_4]$. Therefore, we only need algorithms for finding interaction points a) between the ray and the a 2D line segment and b) between the ray and a triangle. Both approaches are quite standard. They are just briefly summarized here for the convenience of the reader. First, the ray can be defined as
\begin{equation}
	\mb x(t) = \mb x_a + t\left(\mb x - \mb x_a\right).
	\label{eq:def_ray}
\end{equation}
For 2D problems, a point on the line segments $[\mb p_1, \mb p_2]$ can be written as
\begin{equation}
	\mb x(u) = \mb p_1 + u\left(\mb p_2 - \mb p_1\right).
	\label{eq:def_line}
\end{equation}
At the intersection point, we have $\mb x(t) = \mb x(u)$, which yields
\begin{equation}
	\mb x_a + t\left(\mb x - \mb x_a\right) = \mb p_1 + u\left(\mb p_2 - \mb p_1\right)
	\label{eq:line_line_intersect}
\end{equation}
or in the matrix form
\begin{equation}
	\left[\mb x - \mb x_a, \;\mb p_1 - \mb p_2\right]\left[\begin{array}{l} t \\ u\end{array}\right] = \mb p_1 - \mb x_a,
\end{equation}
which can be used for solving the parameter $t$ and $u$. All variables in bold symbols in the above equation are 2D vectors. If the intersection point is inside the line segment, then parameter $u$ must satisfy
\begin{equation}
	0\leq u \leq 1.
\end{equation}
The condition of $t$ for determine $\mb x_0$ and $\mb x_1$ with be discussed later. \\

\begin{figure}[htp]
	\centering
	\subfigure[2D]{\includegraphics[height = .26\textwidth]{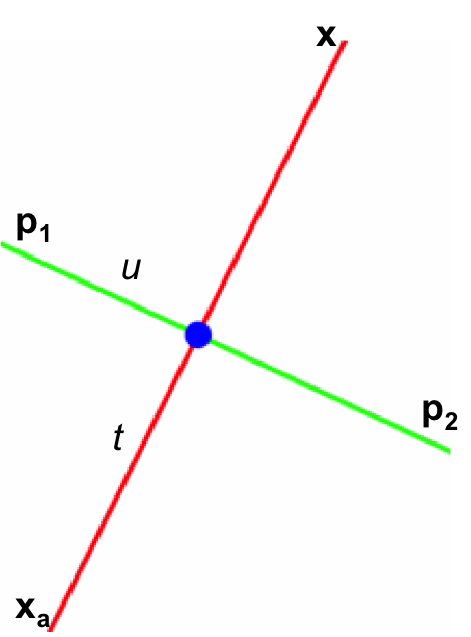}\label{fig:line_line_intersect}}
	\hspace{5pt}
	\subfigure[3D]{\includegraphics[height = .26\textwidth]{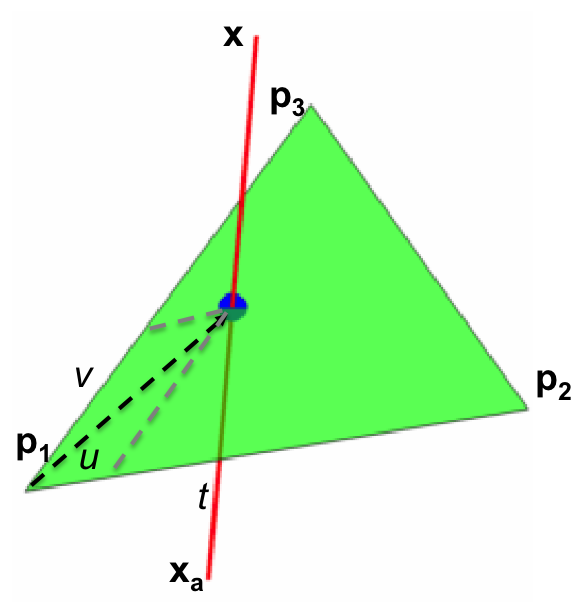}\label{fig:line_surf_intersect}}
	\caption{Intersection of the ray connecting the anchor point $\mb x_a$ and a point in the coupling region $\mb x$, with element on the boundary of the coupling region for 2D case (left) and 3D case (right).}
	\label{fig:intersection_schematics}
\end{figure}

For 3D problems, a point on a surface triangle elements $[\mb p_1, \mb p_2, \mb p_3]$ can be written as
\begin{equation}
	\mb x(u, v) = \mb p_1 + u\left(\mb p_2 - \mb p_1\right) + v\left(\mb p_3 - \mb p_1\right).
	\label{eq:def_surf}
\end{equation}
At the intersection point, $\mb x(t) = \mb x(u, v)$, therefore
\begin{equation}
	\mb x_a + t\left(\mb x - \mb x_a\right) = \mb p_1 + u\left(\mb p_2 - \mb p_1\right) + v\left(\mb p_3 - \mb p_1\right),
\end{equation}
yielding
\begin{equation}
	\left[\mb x - \mb x_a, \;\mb p_1 - \mb p_2, \;\mb p_1 - \mb p_3\right]\left[\begin{array}{l} t \\ u \\ v\end{array}\right] = \mb p_1 - \mb x_a,
\end{equation}
and for a point inside the triangle, the parameters $(u, v)$ must satisfy
\begin{equation}
	0\leq u \leq 1,\;0\leq v \leq 1\;\hbox{and}\;u + v \leq 1.
\end{equation}
When $u+v = 1$, the point is on the edge $[\mb p_2, \mb p_3]$. All the bolded symbols are 3D vectors. The schematics for the ray-segment intersection and ray-triangle intersection are shown in Fig.~\ref{fig:intersection_schematics}.

\subsection{Determine the Intersection Points}
\label{sec:intersect_pt}
In the first part of the section, we introduced the method to determine the elements on the boundary of the coupling region. The basic surface elements for 2D problems are linear bars and for 3D problems are triangles and quadrilaterals, while the quadrilaterals can be divided into two triangles. In the second part of the section, we introduced the algorithms to determine the intersection points between the ray, from an anchor point in the MD region to an arbitrary point in the coupling region, and either a segment or a triangle. Now we have the essential tools for determining the boundary points $\mb x_0$ and $\mb x_1$ for calculating the Arlequin coupling coefficient in Eqn.\eqref{eq:alpha_2D_3D}. \\

Depending on the shape of the coupling region, either one or multiple anchor points can be used, but the algorithms for determining $\mb x_0$ and $\mb x_1$, as well as for calculating the Arlequin coupling coefficient, stay the same. For simplicity, we use FE-MD models with simple MD regions, so that one anchor point is necessary, for the introduction of our algorithms.\\

To find $\mb x_0$ and $\mb x_1$ for an arbitrary point in the coupling region, we need to search for the intersection points between the ray, from the anchor point to the arbitrary point, and all surface elements on the coupling boundary. For regular cases, only two points of intersection are found, corresponding to two values of the parameter $t$ in Eqn.\eqref{eq:def_ray}. By observing the equation, we can conclude the first two conditions for $t$: 1) $t > 0$, otherwise the point is not on the ray and 2) a point with smaller $t$ is closer to the anchor point, i.e. closer to the MD region. Therefore, if only two points of intersection are found, then the one with smaller $t$ is $\mb x_0$ and the other is $\mb x_1$. \\

\begin{figure}[htp]
	\centering
	\includegraphics[width = .46\textwidth]{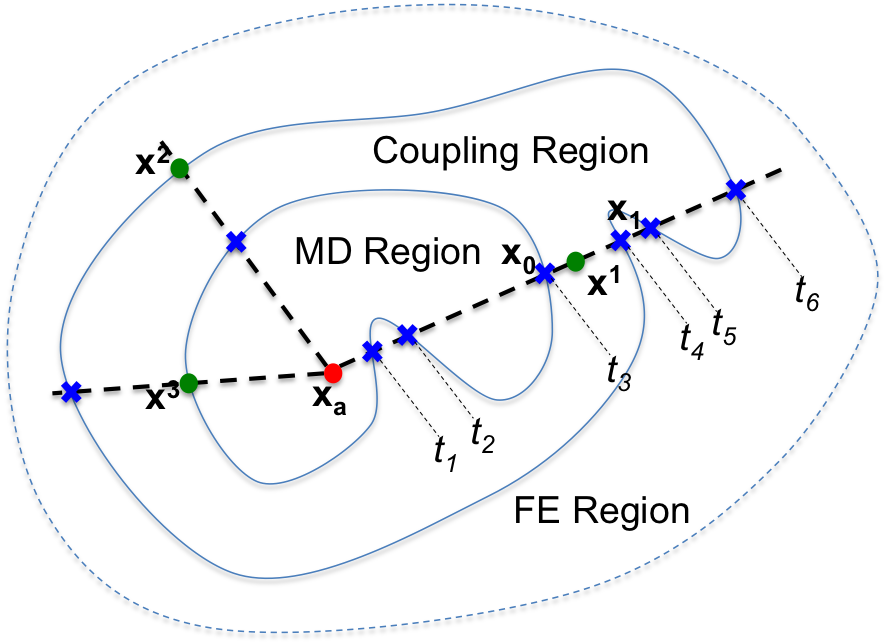}
	\caption{Different cases of the intersection points on the boundary of the coupling region, where $\mb x_a$ is the anchor point, $\mb x^1$, $\mb x^2$ and $\mb x^3$ are some points in or on the coupling region, corresponding to the three types of special cases mentioned in Sec.~\ref{sec:intersect_pt}. $t_1$ to $t_6$ are the values solved from Eqn.\eqref{eq:line_line_intersect} as the candidates of $\mb x_0$ and $\mb x_1$ for $\mb x^1$.}
	\label{fig:alpha_irregular}
\end{figure}

Besides regular cases, there can be three types of irregular cases, as shown in Fig.~\ref{fig:alpha_irregular}. Assuming $\mb x^1$, $\mb x^2$ and $\mb x^3$ are three points in the coupling region, corresponding to the three types of special cases. $\mb x^1$ is inside the coupling region, but its ray has multiple points of intersection with the coupling boundary. $\mb x^2$ is on the FE side of the coupling boundary and $\mb x^3$ is the on the MD side of the coupling boundary. Both $\mb x^2$ and $\mb x^3$ are boundary points themselves. Therefore, from Eqn.\eqref{eq:def_ray}, we have $t > 1$ or $t < 1$ for all the intersection points corresponding to $\mb x^1$, while $t = 1$ for $\mb x^2$ and $\mb x^3$. For $\mb x^1$, by observation, we can see that the proper choice for its $\mb x_0$ is the one with the largest value of $t$ in those with $t < 1$ ($t_3$ in Fig.~\ref{fig:alpha_irregular}), while the proper choice for $\mb x_1$ is the one with the smallest value of $t$ in those $t > 1$ ($t_4$ in Fig.~\ref{fig:alpha_irregular}). For $\mb x^2$, it is $\mb x_1$ by itself, and we only need to apply the above criteria to find $\mb x_0$. For $\mb x^3$, it is $\mb x_0$ by itself, only the criteria for finding $\mb x_1$ is needed. \\

The procedure for finding the boundary points $\mb x_0$ and $\mb x_1$ for an arbitrary point inside the coupling region or on the boundary of the coupling region can be summarized as:
\begin{enumerate}
	\item Identify the surface elements on the coupling boundary based on the mesh of the FE model. If the surface elements are quadrilaterals, break them into triangles;
	\item Choose a proper anchor point inside the MD region;
	\item Find all the points of intersection between the ray defined by Eqn.\eqref{eq:def_ray} and the surface elements on the coupling boundary. The value of $t$ for all valid candidates must be positive, denoted by $\{t_i\}$, with $t_i > 0$;
	\item Denote the value of $t$ for $\mb x_0$ as $T_0$ and that for $\mb x_1$ as $T_1$; Initialize $T_0$ by $0$ and $T_1$ by a large number, denoted by $T_1^{init}$; Then
	\begin{equation}
		\begin{aligned}
			T_0 & = \mathrm{max}\,\left\{t_i | t_i < 1\right\}\\
			T_1 & = \mathrm{min}\,\left\{t_i | t_i  > 1\right\}.
		\end{aligned}
		\label{eq:cond_t}
	\end{equation}
	If $T_0 = 0$, then it means the point inside the coupling region is on the MD side of the coupling boundary. If $T_1 = T_1^{init}$, then it means it is on the FE side. Therefore
	\begin{equation}
		\begin{array}{ll}
			T_0 = 1.0 & \hbox{If $T_0 = 0$ by Eqn.\eqref{eq:cond_t}.}\\
			T_1 = 1.0 & \hbox{If $T_1 = t_1^{init}$ by Eqn.\eqref{eq:cond_t}}.
		\end{array}
		\label{eq:cond_t1}
	\end{equation}
	\item Compute the boundary points $\mb x_0$ and $\mb x_1$ from $T_0$ and $T_1$ by Eqn.\eqref{eq:def_ray}.	
\end{enumerate}
\subsection{Calculating the Coupling Coefficient}
After finding the boundary points $\mb x_0$ and $\mb x_1$, we can then compute the Arlequin coupling coefficient for a point inside the coupling region by Eqn.\eqref{eq:alpha_2D_3D}. For calculating the scaled mass matrix in Eqn.\eqref{eq:def_energies_K} and the scaled stiffness matrix in Eqn.\eqref{eq:def_energies_U} of the FE model, we need to calculate the coefficient for all Gaussian points. For the MD model, the coefficient at the each atomic position in the coupling region need to evaluated. Without adaptivity, such operation needs to be performed only once. Otherwise, this operation needs to be carried out whenever the coupling region changes. In this subsection, we introduce two approaches: the \emph{direct approach} and the \emph{temperature approach}. \\

For the direct approach, the boundary points $\mb x_0$ and $\mb x_1$ are found for every Gaussian points in FE elements and MD atoms in the coupling region, then the Arlequin coupling coefficient is computed by Eqn.\eqref{eq:alpha_2D_3D}. The direct approach can be quite expensive because the number of atoms and Gaussian points in the coupling region can be quite large, and the procedure for searching the boundary points needs to be carried out for each of them. The computational cost is proportional to $(N_{md} + N_{gp})N_{el}^{surf}$, where $N_{md}$ is the number of atoms in the coupling region, $N_{gp}$ is the number of Gaussian points and $N_{el}^{surf}$ is the number of surface elements on the coupling boundary. \\

\begin{figure}[htp]
	\centering
	\subfigure[Direct Approach]{\includegraphics[width = .3\textwidth]{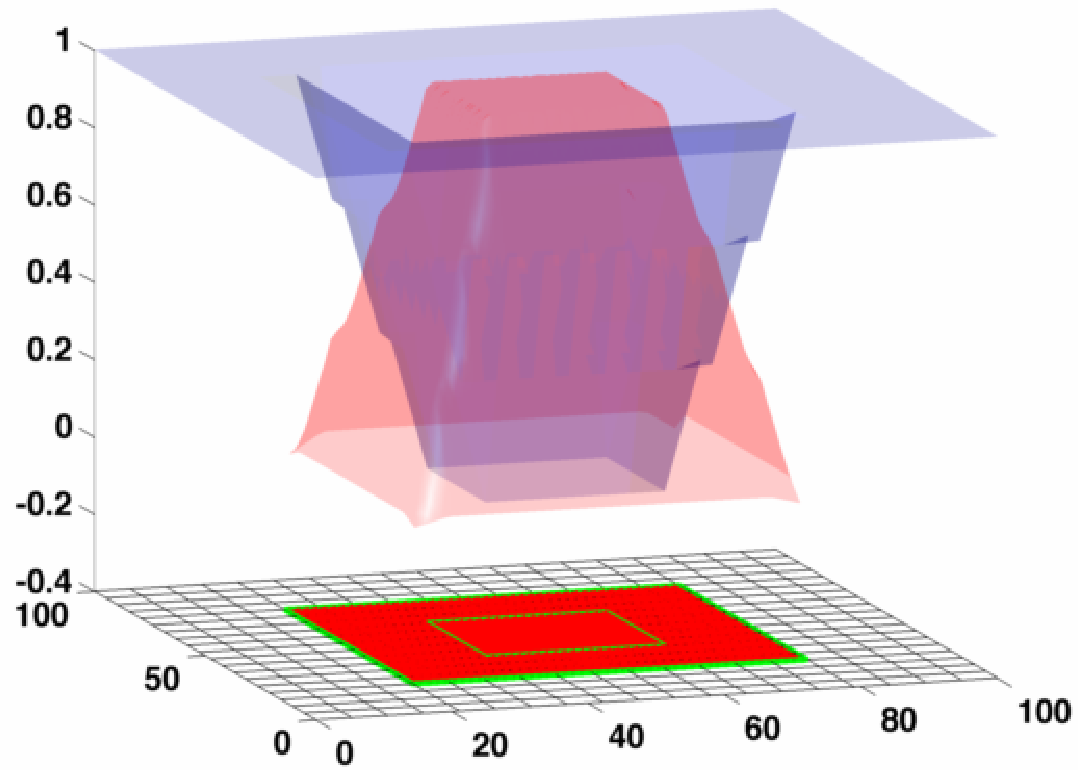}}\\
	\subfigure[Temperature Approach]{\includegraphics[width = .3\textwidth]{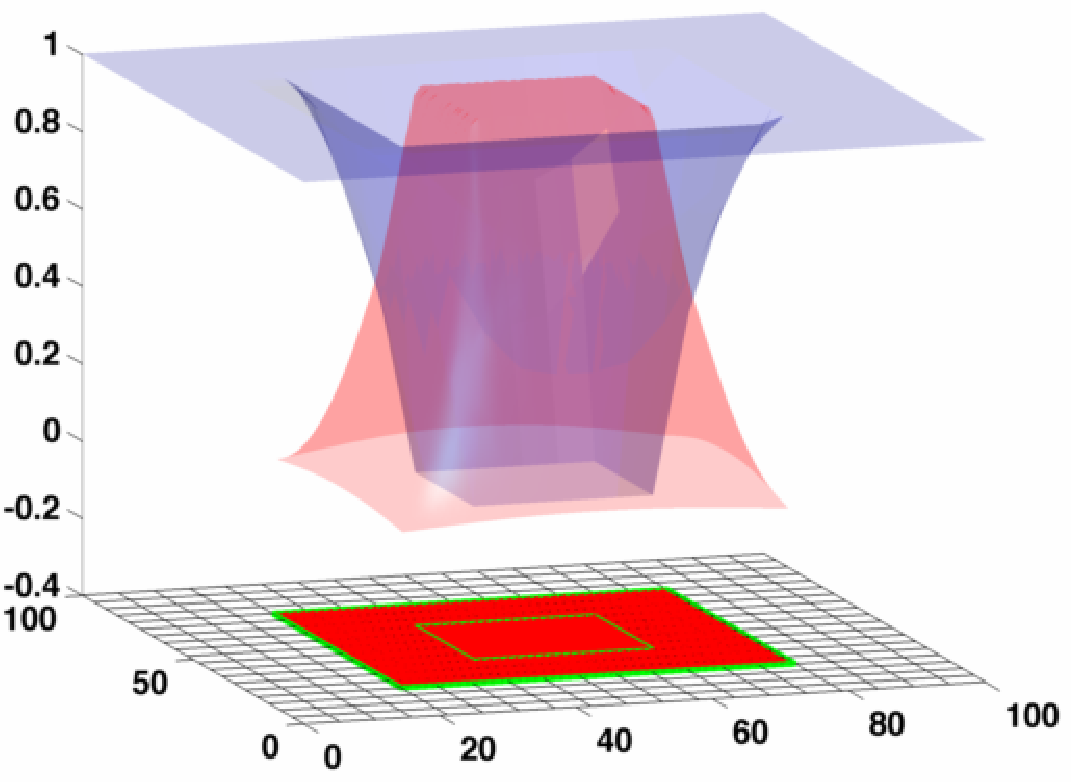}}
	\caption{Examples of the computed Arlequin coupling coefficient for 2D problems by the direct approach (top) and by the temperature approach (button), where the MD region is in the middle and the coupling boundary is highlighted by green. The coefficient $\alpha$ is used as the scaling factor for the energies of FE model in the coupling region, while $1-\alpha$ as the scaling factor for the MD model.}
	\label{fig:eg_alpha_2D}
\end{figure}

\begin{figure}[htp]
	\centering
	\subfigure[Direct Approach, MD]{\includegraphics[width = .3\textwidth]{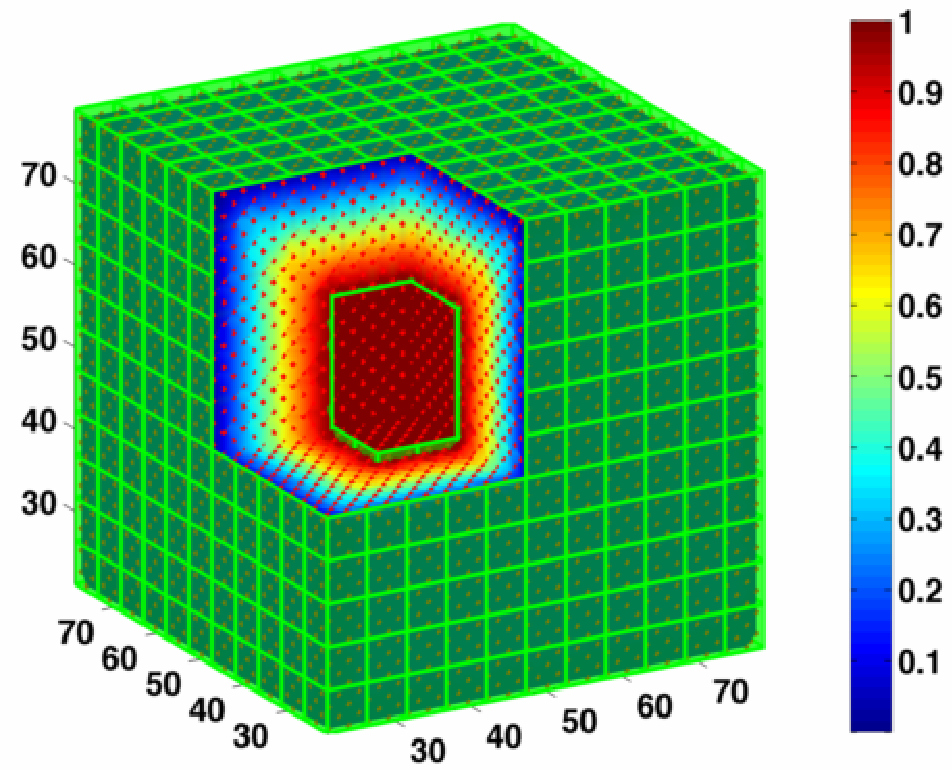}}\\
	\subfigure[Direct Approach, FE]{\includegraphics[width = .3\textwidth]{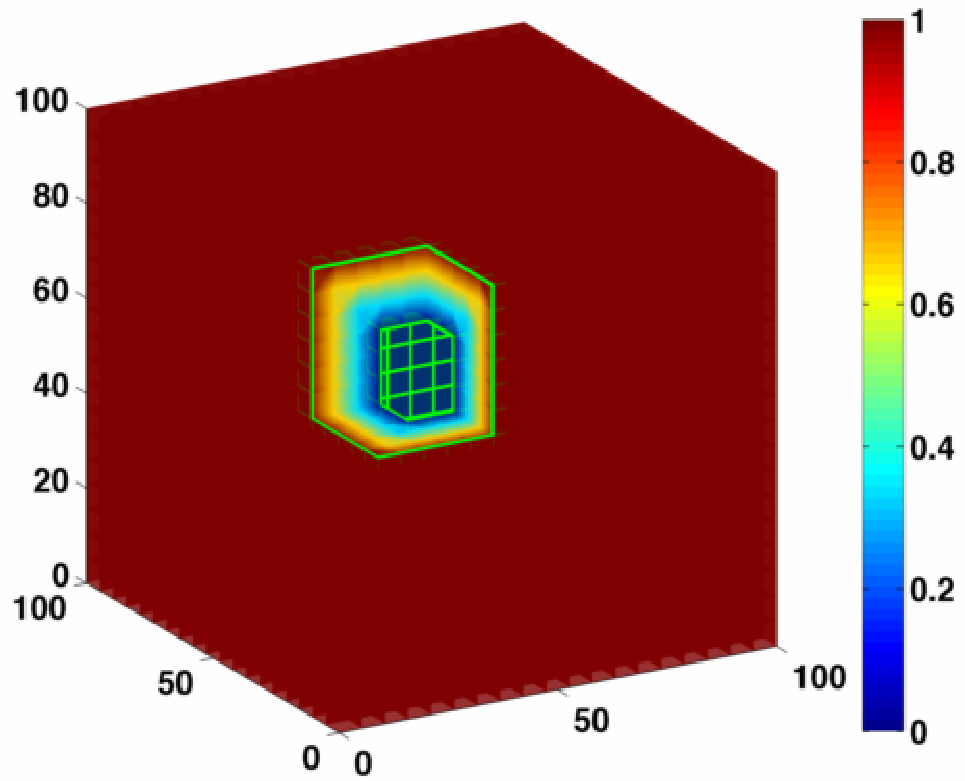}}\\
	\subfigure[Temperature Approach, MD]{\includegraphics[width = .3\textwidth]{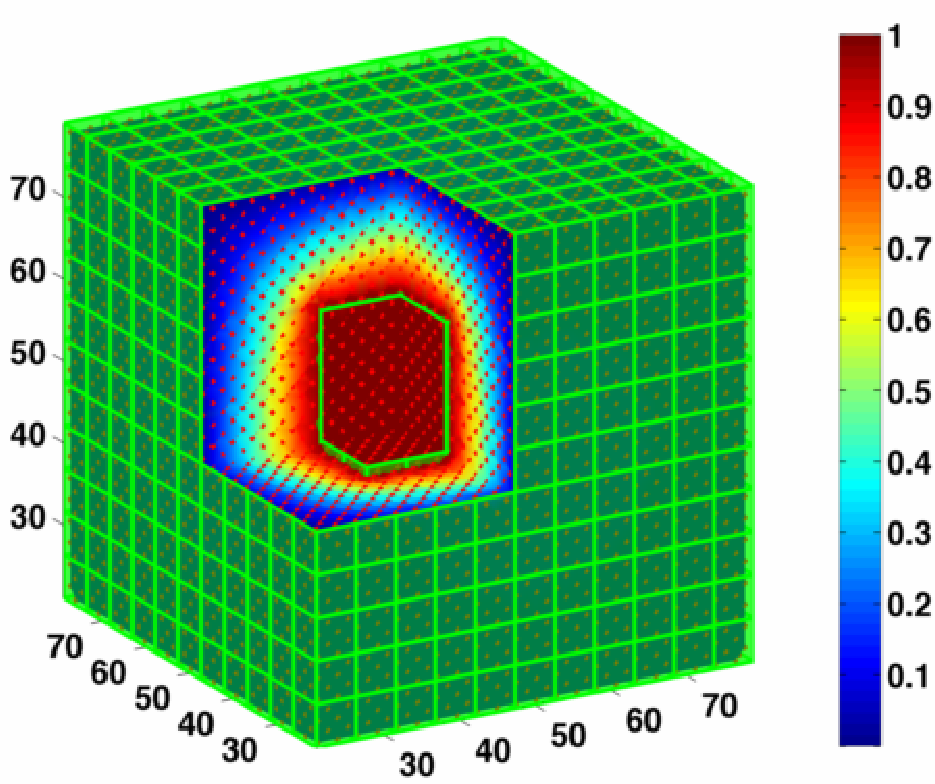}}\\
	\subfigure[Temperature Approach, FE]{\includegraphics[width = .3\textwidth]{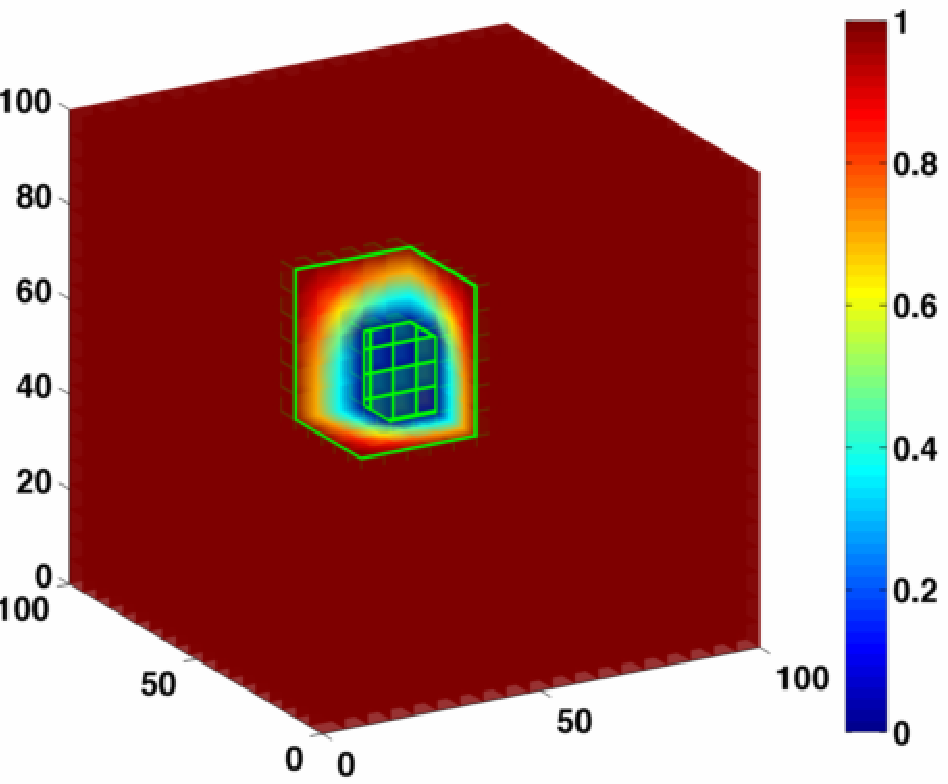}}
	\caption{Examples of the computed Arlequin coupling coefficient for 3D problems from the direct approach, (a) and (b), and from the temperature approach, (c) and (d), where the MD region is inside the middle and the coupling boundary is highlighted by green. The color represents the value of the scaling factor, $\alpha$ for the FE model and $1-\alpha$ for the MD model.}
	\label{fig:eg_alpha_3D}
\end{figure}

Here we introduce a more efficient approach: the \emph{temperature approach}. For this approach, we consider the Arlequin coupling coefficient as temperature, as both of them are positive scalars. Since the coefficient is zero on the MD side of the coupling boundary and one on the FE side. We can use them as the boundary conditions. Then the coefficient for all the FE nodes inside the coupling region can be computed by solving a linear thermal conduction problem, i.e.
\begin{equation}
	\begin{aligned}
		&\mb K_{\theta}\hat{\tens\alpha}= \left(\intV{\Omega_{cp}}{\mb B_{\theta}^T\tens\kappa_{\alpha}\mb B_{\theta}}\right)\hat{\tens\alpha} = \mb 0\\
		&\alpha_0 = 0\hspace{10pt}\hbox{and}\hspace{10pt}\alpha_1 = 1,
	\end{aligned}	
	\label{eq:alpha_therm}
\end{equation}
where $\mb B_{\theta}$ is the temperature-gradient matrix \cite{wriggers2008}, $\tens\kappa_{\alpha}$ is the conductivity matrix which can be simply set to the identity matrix. $\hat{\tens\alpha}$ is the vector containing the values of the coefficient at FE nodes in the coupling region, $\alpha_0$ means the value of the coefficient at the MD side of the coupling boundary and $\alpha_1$ the value at the FE side of the coupling boundary. After solving for the nodal values, its value at the Gaussian points in the FE elements can be interpolated by the shape functions. On the other hand, the iso-parametric coordinates of all the atoms in the coupling region are obtained by the inverse iso-parametric mappings described in Sec.~\ref{sec:inv_iso_coord}. Therefore, the Arlequin coupling coefficient for the atoms can be also interpolated from the nodal values by the FE shape functions. However, the coefficient obtained by the temperature approach is slightly nonlinear. 2D and 3D examples of the coupling coefficient obtained by these two approaches are given in Fig.~\ref{fig:eg_alpha_2D} and Fig.~\ref{fig:eg_alpha_3D}.\\

After generating the mesh for the FE model and creating atoms for the MD model, the remaining steps of the preprocessing for a concurrent FE-MD model can be summarized as
\begin{enumerate}
	\item Perform the inverse iso-parametric mapping introduced in Sec.~\ref{sec:inv_iso_coord} on all the atoms, with respect to the FE elements; This step not only identifies the coupling region, but also determines the iso-parametric coordinates of the atoms in the coupling region inside the FE elements, which will be used later;
	\item Identify the surface elements on the coupling boundary by the procedure introduced in Sec.~\ref{sec:ix_bc};
	\item Calculate the Arlequin coupling coefficient $\alpha$ for all the FE nodes in the coupling region:
	\begin{itemize}
		\item Choose proper anchor point inside the MD region, according to its geometry;
		\item If the direct approach is chosen, then search for the boundary points $\mb x_0$ and $\mb x_1$ for all FE nodes in the coupling region by the procedure introduced in this section and compute $\alpha$ according to Eqn.\eqref{eq:alpha_2D_3D};
		\item If the temperature approach is chosen, it is only necessary to identify the FE nodes on the boundary of the coupling region, by using the boundary ray-tracing algorithm. Then, the nodal values of $\alpha$ inside the coupling region can be determined by the thermal conduction analogy, Eqn.\eqref{eq:alpha_therm}.
	\end{itemize}
	\item Determine the values of $\alpha$ at all the Gaussian points in the FE elements in the coupling region by interpolating the nodal values;
	\item Determine the values of $\alpha$ at all atomic positions in the coupling region also by interpolating the nodal values of the FE elements containing them. In this step, the iso-parametric coordinates obtained at the first step are necessary. Once the values of $\alpha$ is known, the scaling factor used for the MD model is $1 - \alpha$, as shown in Fig.~\ref{fig:eg_alpha_2D} and Fig.~\ref{fig:eg_alpha_3D}.
\end{enumerate}
\section{Numerical Examples}
\label{sec:example}
\begin{figure}[htp]
	\centering
	\includegraphics[width = .3\textwidth]{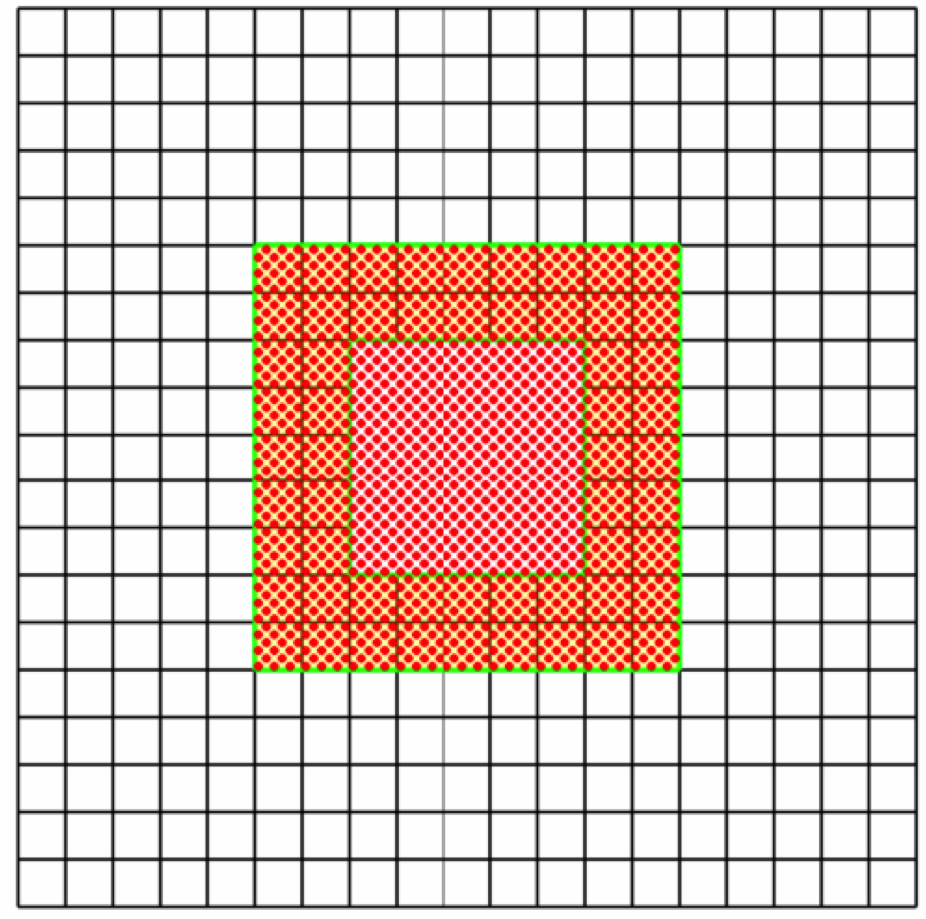}
	\caption{Initial configuration of the 2D example.}
	\label{fig:eg_model}
\end{figure}

In this section, we use a 2D example to demonstrate the importance of Arlequin coupling coefficient. The initial configuration of the model is shown in Fig.~\ref{fig:eg_model}. A Lennard-Jones type of potential is used
\begin{equation}
	\phi(r) = \epsilon\left[\left(\frac{n}{m}\right)\left(\frac{r_0}{r}\right)^m - \left(\frac{r_0}{r}\right)^n\right],
\end{equation}
where $r_0$ is the equilibrium distance. Since the actual physical units are of no importance in this work, the following parameters are used:
\begin{equation}
	\nonumber
	\epsilon = 1,\,\,n = 6,\,\,m = 12\,\,\hbox{and}\,\,r_0 = 1.2405.
\end{equation}
The MD model is a 2D lattice with lattice vectors:
\begin{equation}
	\nonumber
	\mb a_1 = r_0\left[\frac{\sqrt{2}}{2}, -\frac{\sqrt{2}}{2}\right]^T\hspace{5pt}\hbox{and}\hspace{5pt}\mb a_2 = r_0\left[\frac{\sqrt{2}}{2}, \frac{\sqrt{2}}{2}\right]^T.
\end{equation}
The FE model has uniform mesh with linear quadrilateral elements. The length of the edge of each element is $5.2632$. There are 336 elements in the FE model and $1458$ atoms in the MD model. Among them, 56 elements and 1008 elements are identified in the coupling region. In this numerical example, only linear calculation is performed. The coupled model is shown in Fig.~\ref{fig:eg_model}.\\

\begin{figure}[htp]
	\centering
	\includegraphics[width = .2\textwidth]{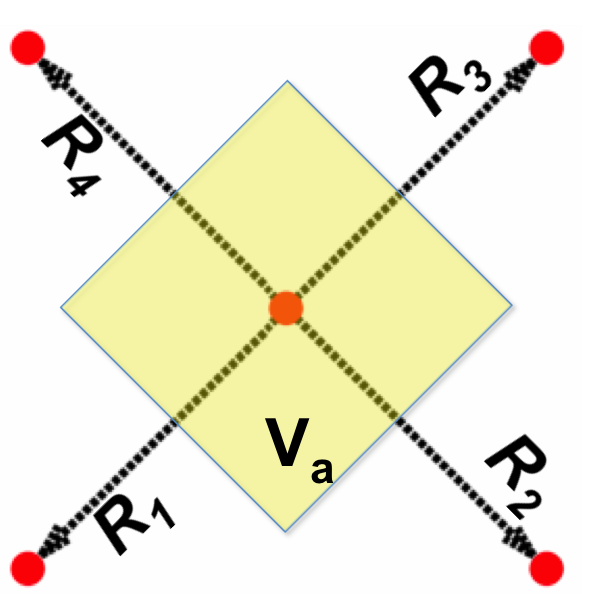}
	\caption{Representative lattice for the 2D FE model in Sec.~\ref{sec:example}, for the calculation of the elastic tensor. Shaded area is the Wigner-Seitz cell of the lattice.}
	\label{fig:eg_rep_lat}
\end{figure}

The unscaled tangent matrix for the MD model is
\begin{equation}
	\nonumber
	\mb K_q = \mathrm{Assemb}\,\mb K_{km},
\end{equation}
where $\mathrm{Assemb}\,\mb K_{km}$ means assembling the local tangent matrices $\mb K_{km}$ into the global one $\mb K_q$. $k$ and $m$ are indices of atoms. $\mb K_{km}$ is the 2nd order derivative of the MD potential energy at the initial configuration with respect to the displacements of atom $k$ and  atom $m$, i.e.
\begin{equation}
	\nonumber
	\begin{aligned}
		& \mb K_{km} = \left.\pd{^2\mathcal U^{MD}}{\mb q_k\partial\mb q_m}\right |_{\hat{\mb q} = \mb 0}\\
		& = \inv{2}\sum_i\sum_{n\in\mathcal N(i)}\left[\left(\phi''_{in}\inv{R_{in}^2} - \phi'_{in}\inv{R_{in}^3}\right)\mb R_{in}\otimes\mb R_{in} + ...\right.\\
		& \left. + \phi'_{in}\inv{R_{in}}\mb I\right]\left(\delta_{nk} - \delta_{ik}\right)\left(\delta_{nm} - \delta_{im}\right)
	\end{aligned}
\end{equation}
where $\mb R_{in}$ is the distance vector from atom $i$ to atom $n$ in the initial configuration and $R_{in}$ is its magnitude. $\mathcal N(i)$ denotes the set formed by the interaction neighbors of atom $i$. $\phi'$ and $\phi''$ are the first and the second order derivatives of the potential evaluated at $R_{in}$. $\delta_{...}$ is the Kronecker-delta. For detailed derivation, one can refer to \cite{shan2009d}. The elastic tensor for the FE model can be derived as \cite{shan2009d}:
\begin{equation}
	\nonumber
	\mathbb C = \frac{1}{2V_a}\sum_{j = 1}^4 \left(\frac{\phi''_j}{R_j^2} - \frac{\phi'}{R_j^3}\right)\mb R_j \otimes\mb R_j\otimes\mb R_j\otimes\mb R_j
\end{equation}
where $\mb R_j$ are the lattice vectors in the representative lattice:
\begin{equation}
	\nonumber
	\begin{array}{ll}
		\mb R_1 = \frac{r_0}{\sqrt{2}}\left[-1, -1\right]^T, & \mb R_2 = \frac{r_0}{\sqrt{2}}\left[\;1, -1\right]^T\\
		\mb R_3 = \frac{r_0}{\sqrt{2}}\left[\;\;\,1, \;\;\,1\right]^T, & \mb R_4 = \frac{r_0}{\sqrt{2}}\left[-1, \;1\right]^T
	\end{array}
\end{equation}
and $V_a = r_0^2/2$ is the volume of the Wigner-Seitz cell of the lattice. The representative lattice is plotted in Fig.~\ref{fig:eg_rep_lat}. More details about the representative lattice can be found in \cite{shan2009d}.\\

The focus of this work is on the calculation of the coupling coefficient. We perform only linear calculation on the model. Therefore, the tangent matrices are only computed once at the initial configuration. To make sure that the system is at equilibrium at the initial configuration, only the nearest neighbor for the atomic interaction is considered. This is also the reason that the representative lattice for the FE model is so simple, as shown in Fig.~\ref{fig:eg_rep_lat}.\\
\begin{figure}[htp]
	\centering
	\includegraphics[width = .4\textwidth]{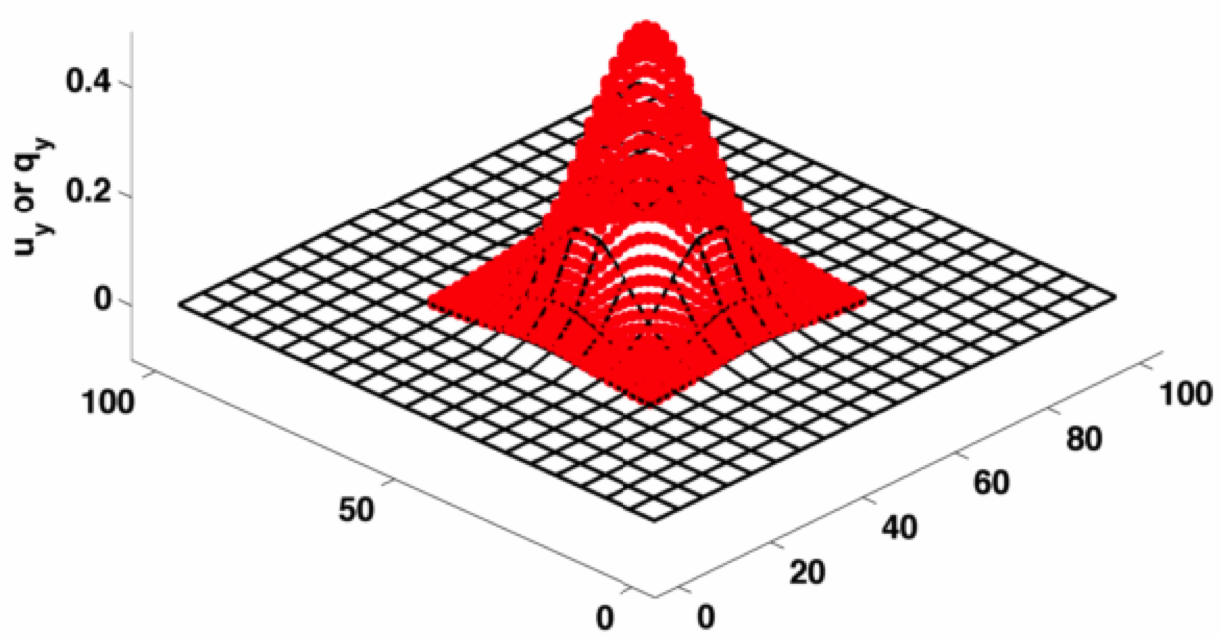}
	\caption{Y-component of the displacement field at the initial step.}
	\label{fig:eg_init}
\end{figure}

\begin{figure}[htp]
	\centering
	\subfigure[Direct Arlequin]{\includegraphics[width = .3\textwidth]{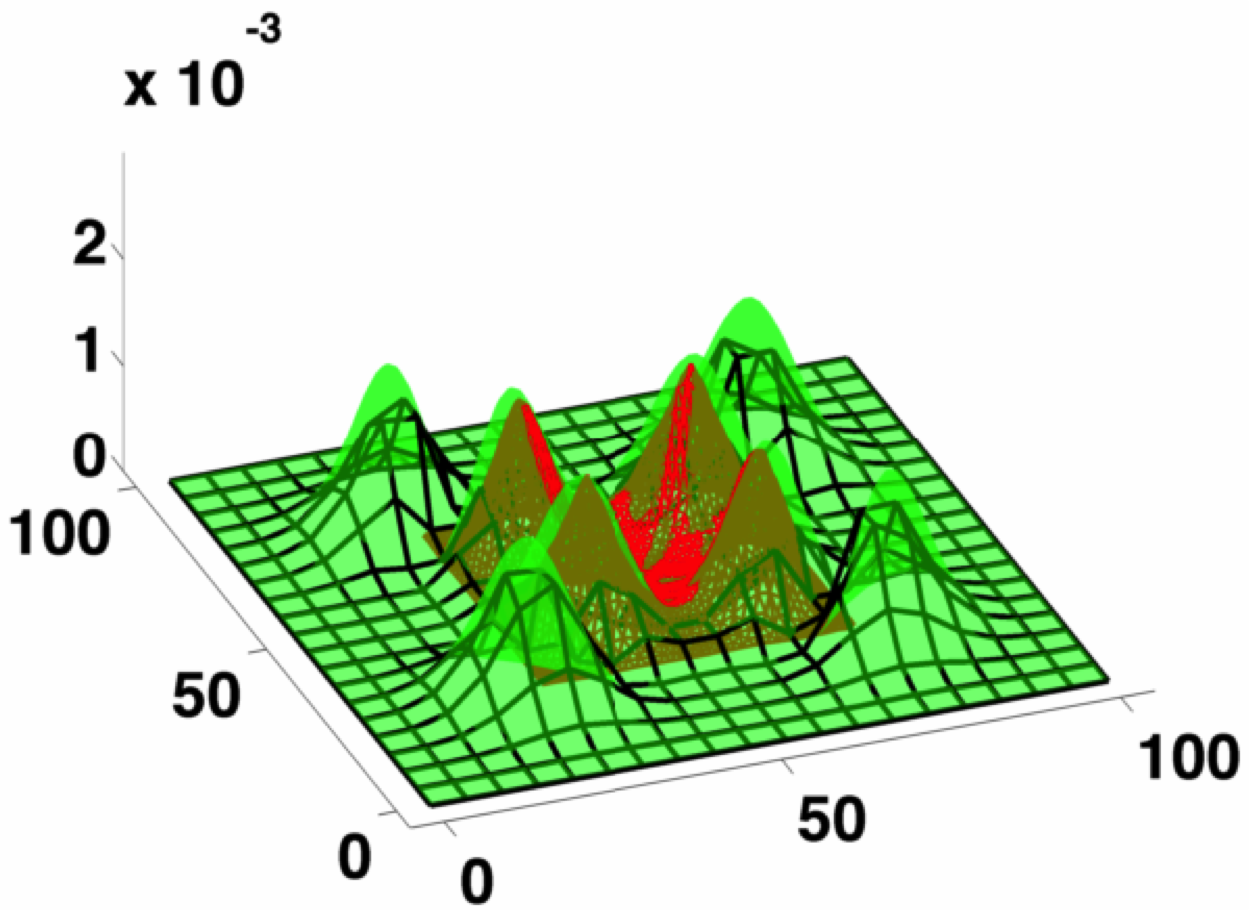}}
	\subfigure[Temperature Arlequin]{\includegraphics[width = .3\textwidth]{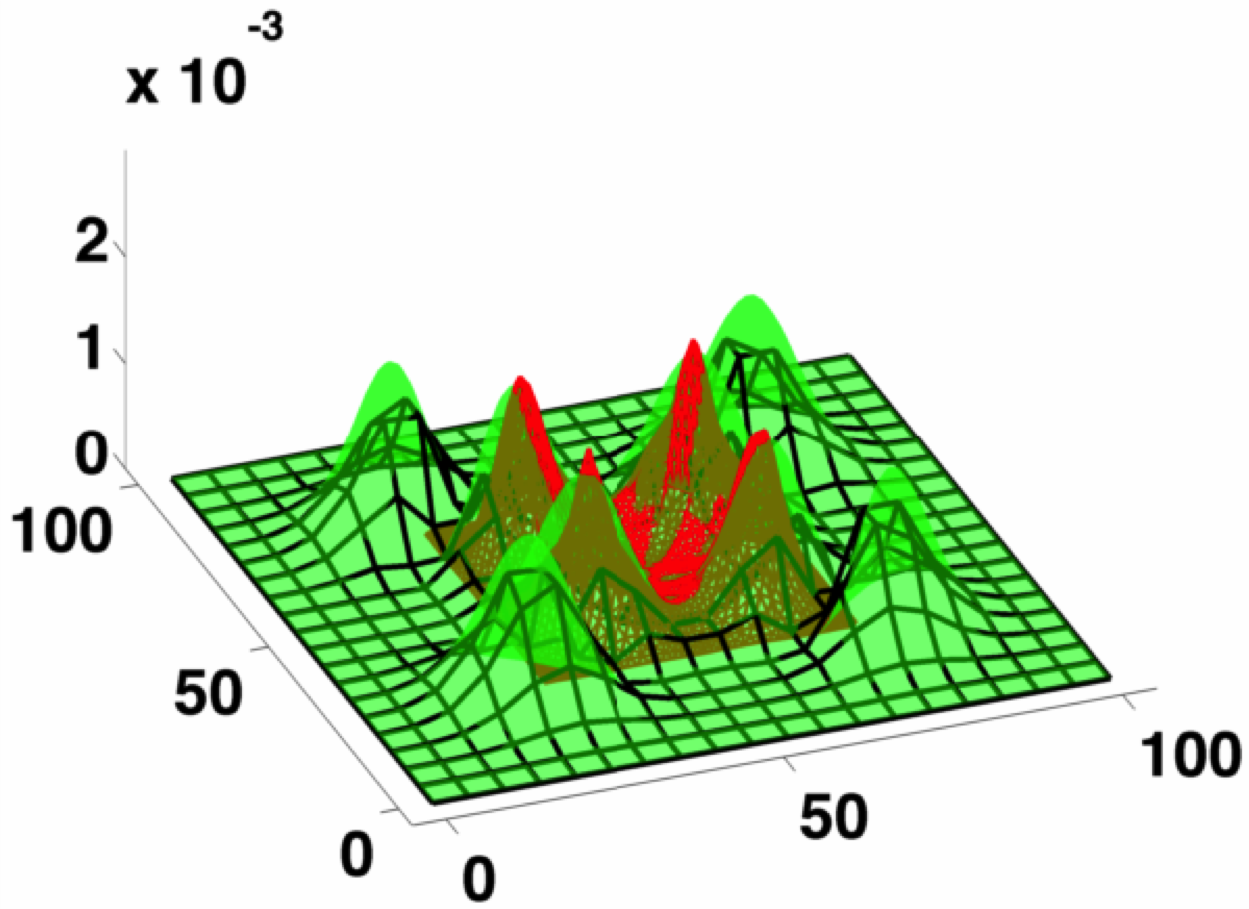}}
	\subfigure[No Scaling]{\includegraphics[width = .3\textwidth]{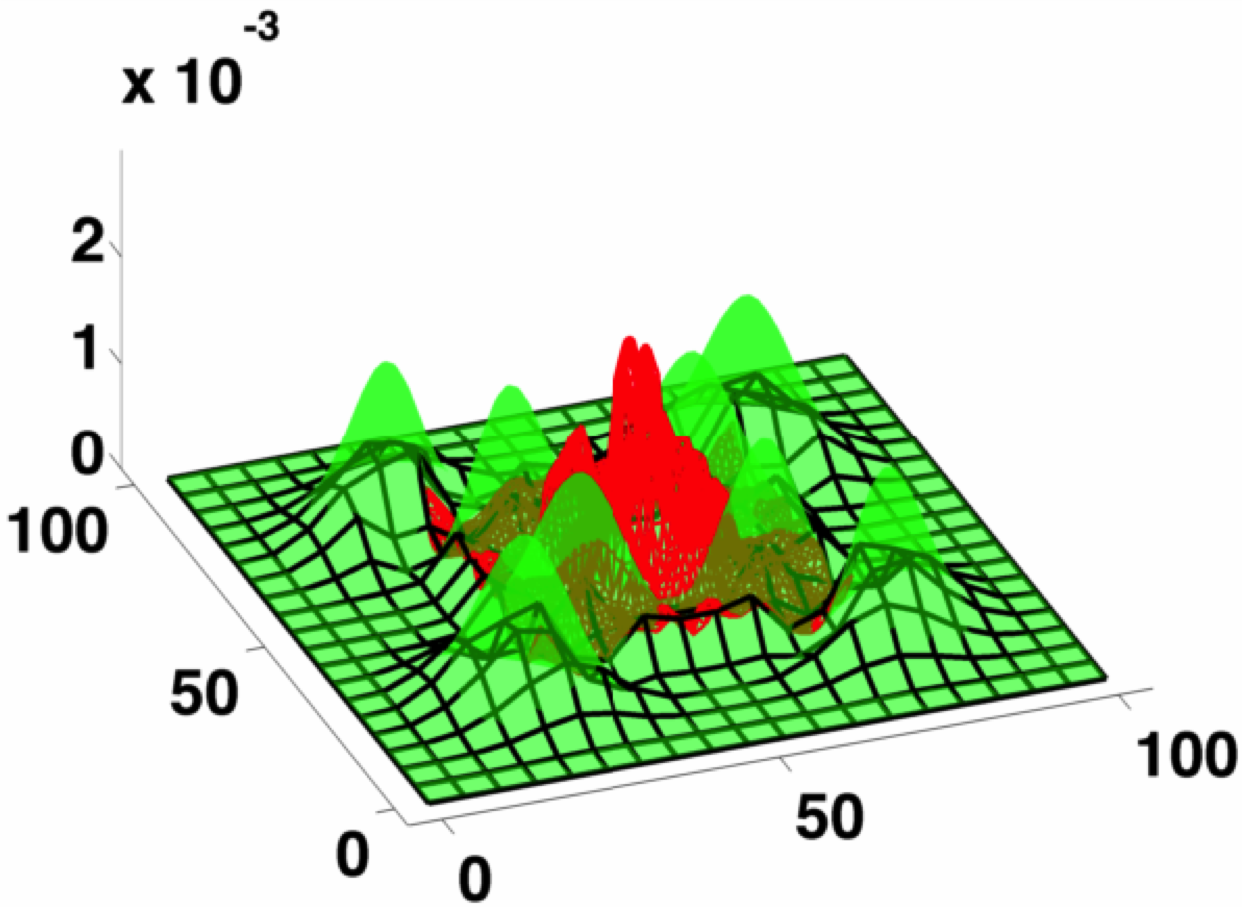}}
	\subfigure[Constant Scaling]{\includegraphics[width = .3\textwidth]{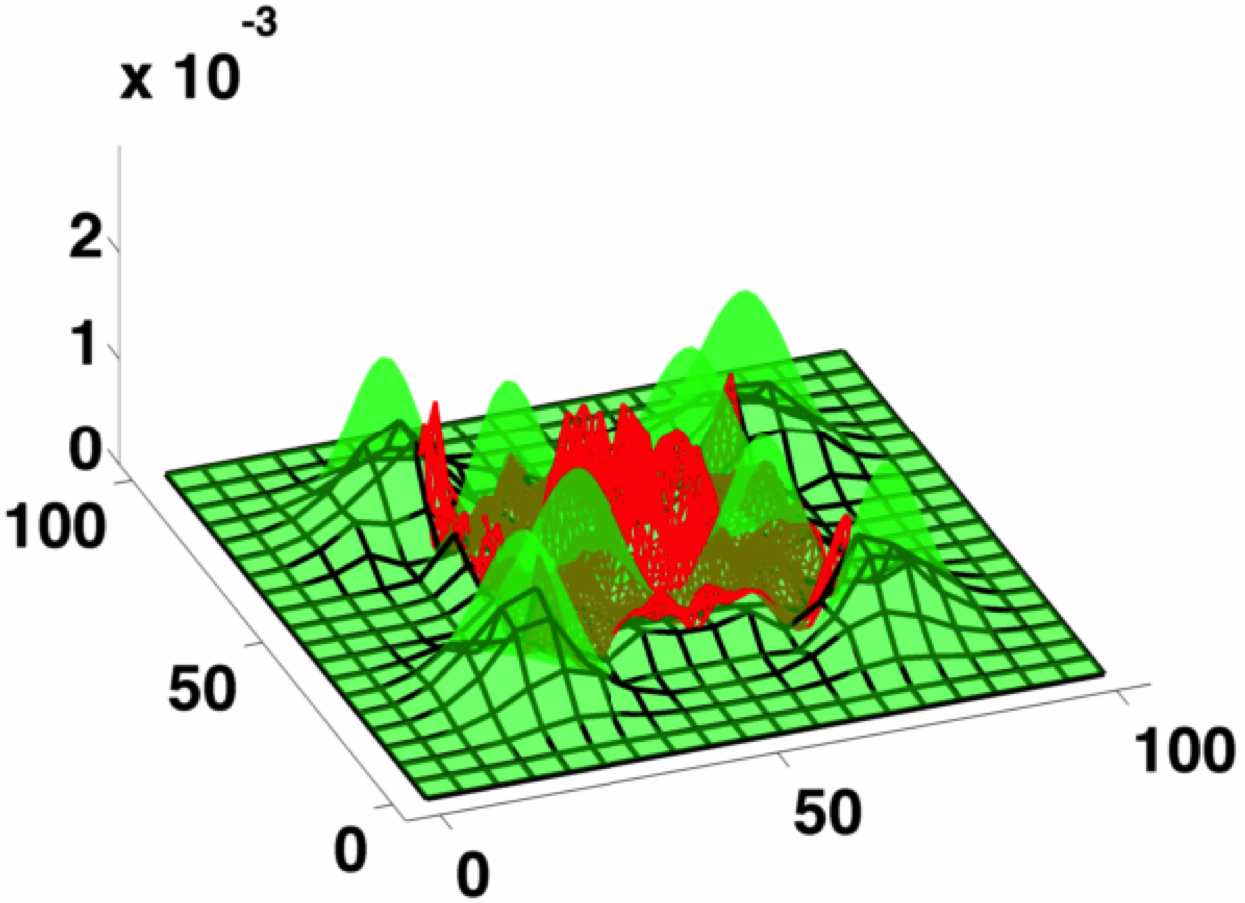}}
	\caption{Snapshot at the 120th step for the propagation of \textbf{kinetic energy density} for FE-MD models with different energy scaling approaches in the coupling region. The result of the full MD model is plotted as a green transparent membrane overlaid on the result of the FE-MD model.}
	\label{fig:snapshot_120}
\end{figure}

\begin{figure}[htp]
	\centering
	\subfigure[Direct Arlequin]{\includegraphics[width = .3\textwidth]{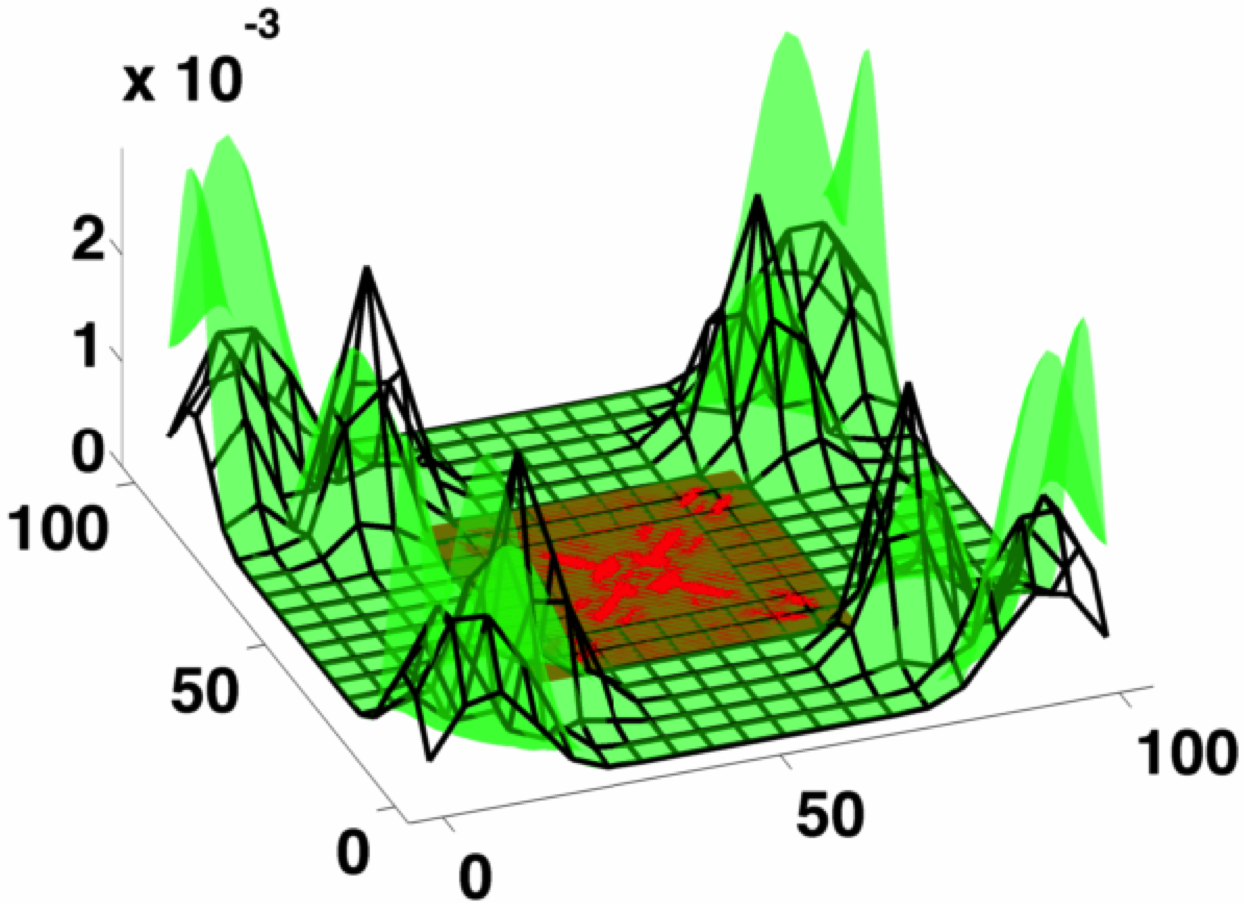}}
	\subfigure[Temperature Arlequin]{\includegraphics[width = .3\textwidth]{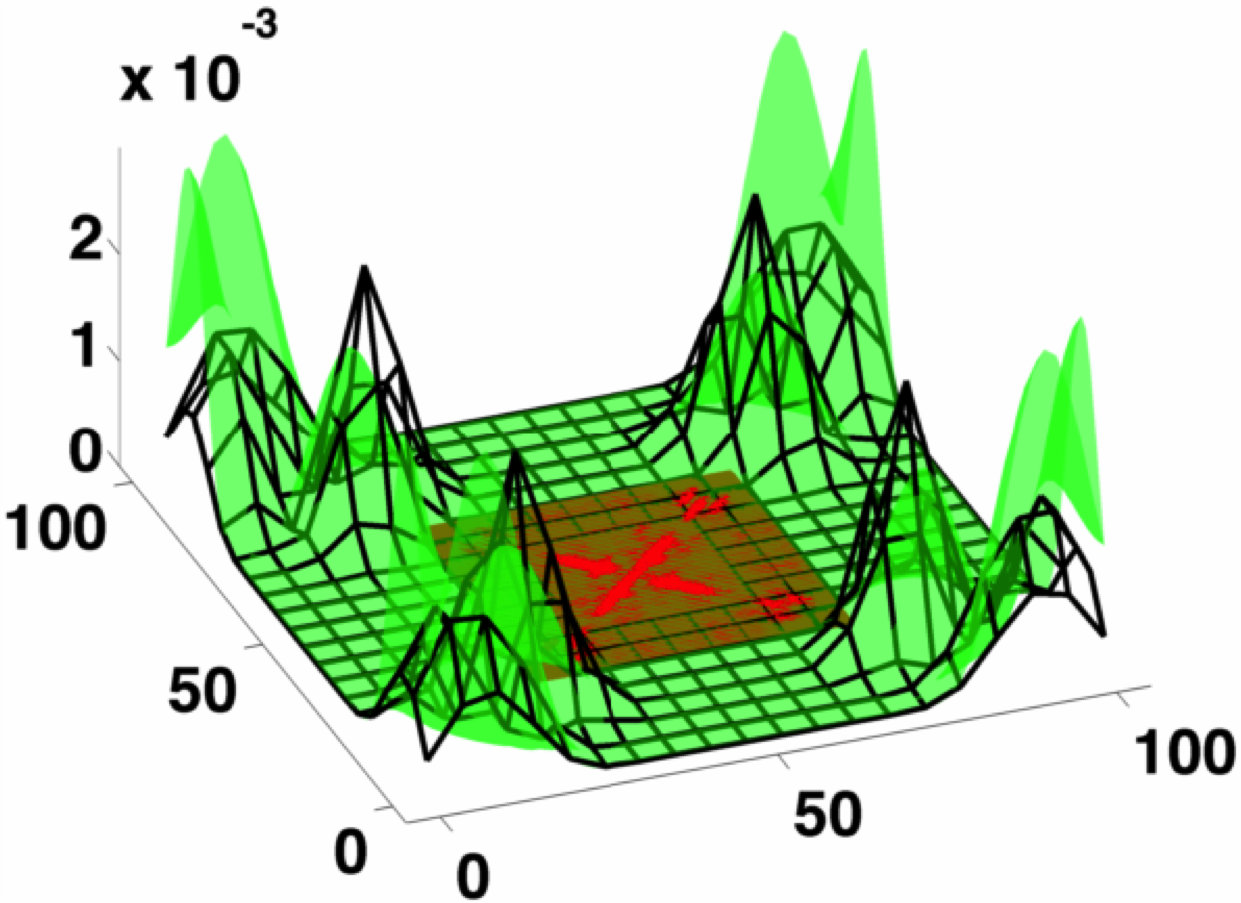}}
	\subfigure[No Scaling]{\includegraphics[width = .3\textwidth]{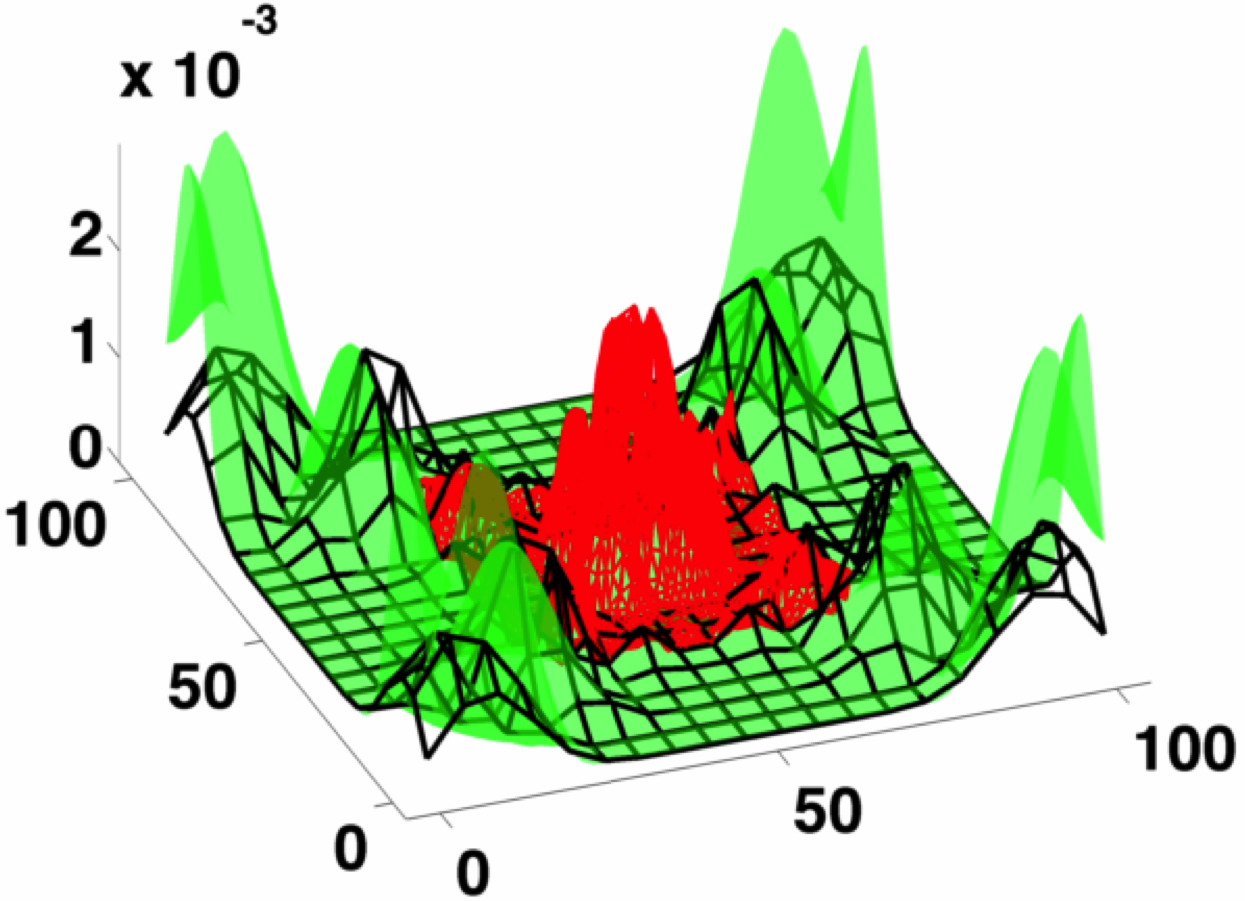}}
	\subfigure[Constant Scaling]{\includegraphics[width = .3\textwidth]{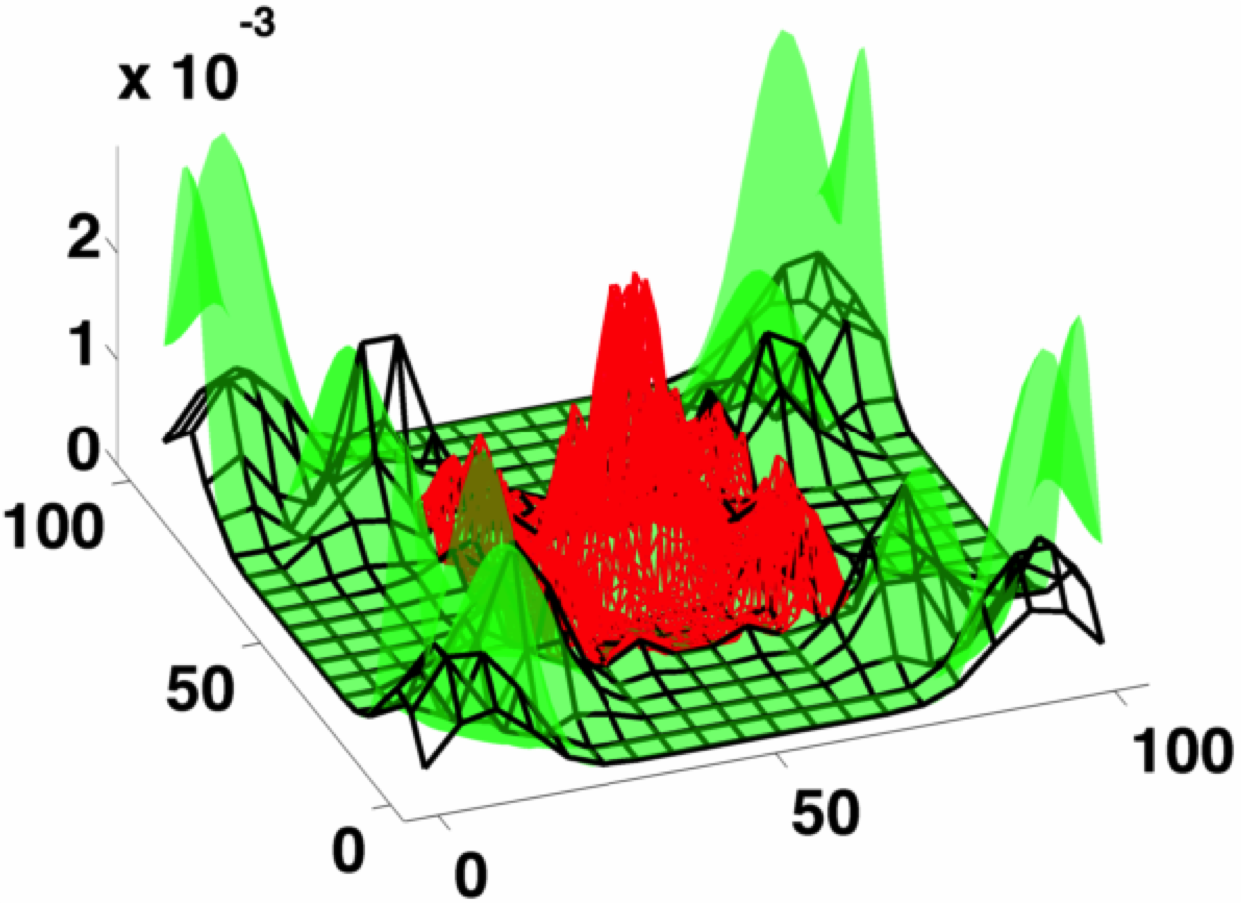}}
	\caption{Snapshot at the 220th step for the propagation of \textbf{kinetic energy density} for FE-MD models with different energy scaling approaches in the coupling region. The result of the full MD model is plotted as a green transparent membrane overlaid on the result of the FE-MD model.}
	\label{fig:snapshot_220}
\end{figure}

The weak coupling method is used for this example, and the equations of motion Eqn.\eqref{eq:eom_general} for the model are integrated explicitly, by the central difference method \cite{wriggers2008} with a time step size of $0.04$. The y-component of the displacement field are initialized as a Gaussian function, as shown in Fig.~\ref{fig:eg_init}. The x-component of the displacement field, velocity and acceleration are also set to zero at the initial step.  A full-scale MD model with same material and geometry parameters are also created. The propagation of densities of the kinetic energy, potential energy of the coupled FE-MD model is then compared with these of the full MD model. To investigate the importance of the Arlequin coupling coefficient, we compared the results obtained from:
\begin{enumerate}
	\item FE-MD model with energies of different models in the coupling region scaled by the Arlequin coupling coefficient obtained by the direct approach;
	\item FE-MD model with energies of different models in the coupling region scaled by the Arlequin coupling coefficient obtained by the temperature approach;
	\item FE-MD model with energies of different models in the coupling region not scaled at all;
	\item FE-MD model with energies of different models in the coupling region scaled by a constant ($0.5$), so that Eqn.\eqref{eq:def_coupling_energy} becomes:
		\begin{equation}			
			\Pi_{cp} = 0.5\Pi^{FE}_{cp} + 0.5\Pi^{MD}_{cp}.
			\label{eq:def_alpha_half}
		\end{equation}	
\end{enumerate}
Snapshots of the propagation of the kinetic energy density for the above four types of model are plotted in Fig.~\ref{fig:snapshot_120} and Fig.~\ref{fig:snapshot_220}. The former is captured at the 120th step and the later at 220th step. From Fig.~\ref{fig:snapshot_120}, we can observe that when the Arlequin coupling coefficient $\alpha$ is not used, significant amount of kinetic energy is reflected back into the MD region from the coupling region, despite that all models are initialized by low-frequency motions, which implies that artificial wave impedance will be introduced in the coupling region, even for low-frequency motion which CAN be passed into the FE model from the MD model. On the other hand, the difference between the behavior of the model with $\alpha$ obtained from the direct approach and from the temperature approach is much less insignificant. Both models match the full-MD model quite well. From Fig.~\ref{fig:snapshot_220}, we can see that all kinetic energy can be transferred into the FE model for the models when the Arlequin coupling coefficient is used. The artificial reflection of energy due to improper choice of the scaling factors can be more clearly observed, as the reflected part of the energy in Fig.~\ref{fig:snapshot_120} appears to be trapped inside the MD region in Fig.~\ref{fig:snapshot_220}. In this figure, one can also observe that the inconsistency between the FE-MD models and the full-MD model is significant at the corners. This is due to the modeling error in the computation of the FE elastic tensor, which uses potential energy density of an internal atom while the full-MD model uses open boundary, instead of periodic boundary conditions. Since the focus of this work is the computation of the coupling coefficient for the Arlequin approach, we do not use special treatment for the FE elements on the open boundary, in order to simplify the implementation. For the treatment of the FE elements on the open boundary, pleaser refer to our previous work \cite{shan2009d}. \\

\begin{figure}[htp]
	\centering
	\subfigure[Kinetic Energy]{\includegraphics[width = .4\textwidth]{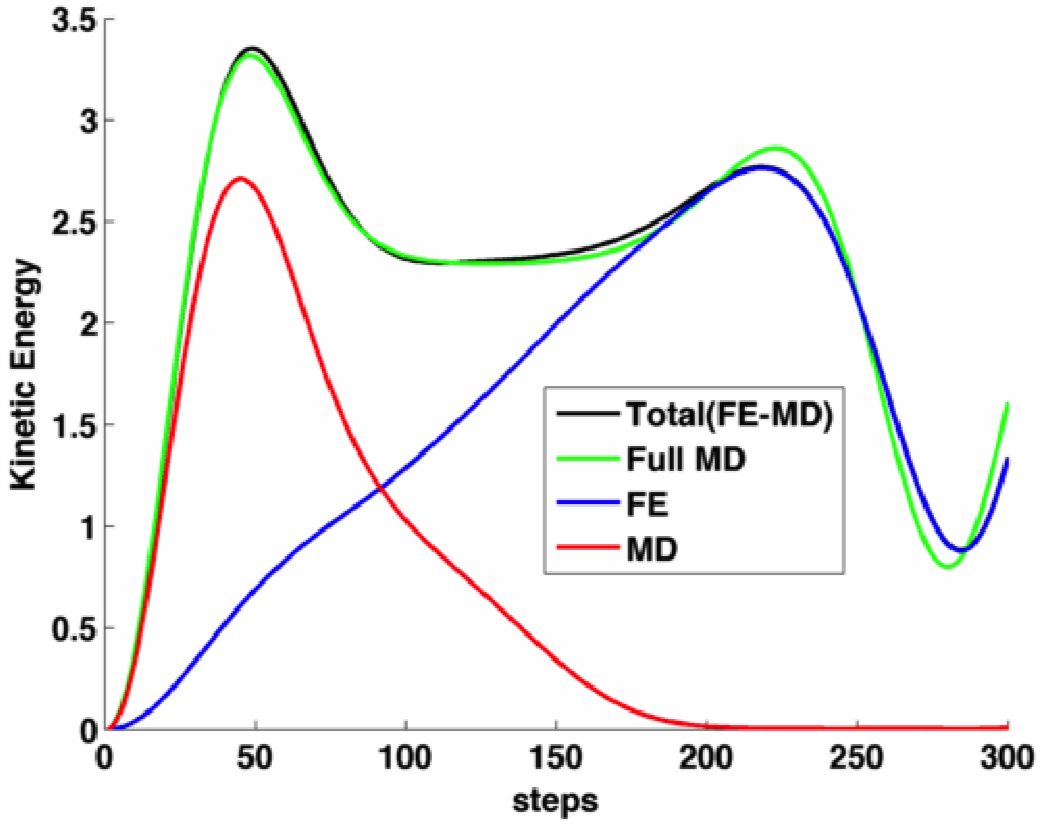}}
	\subfigure[Potential Energy]{\includegraphics[width = .4\textwidth]{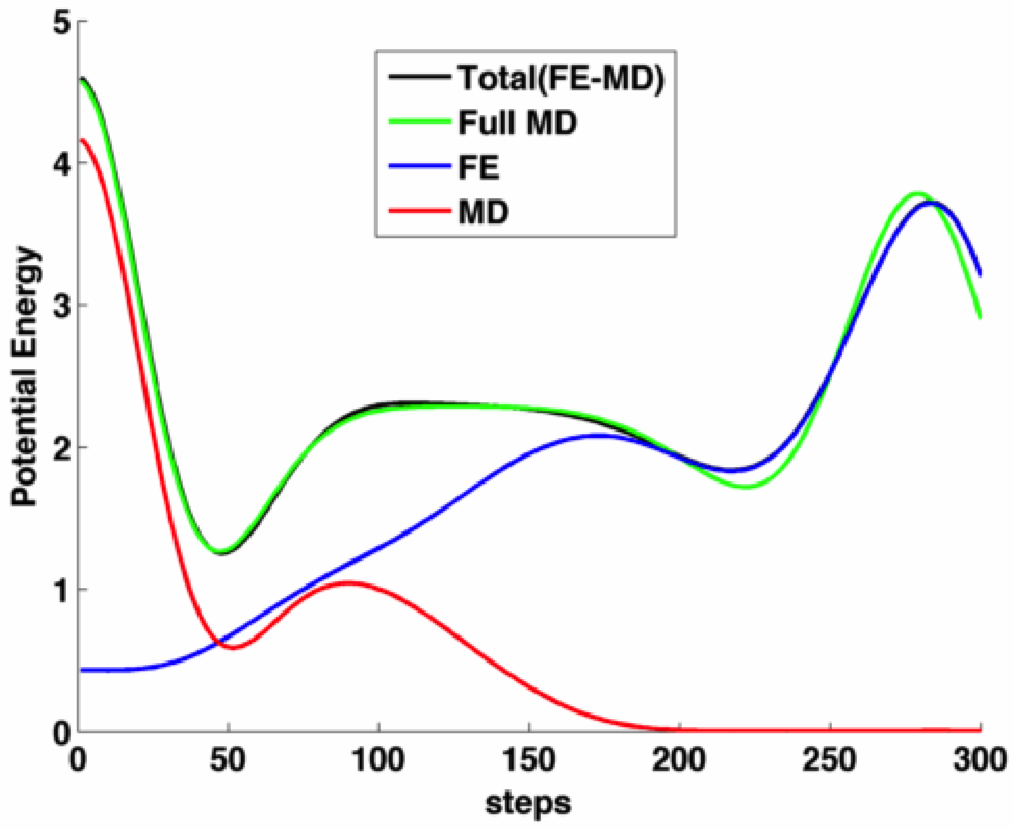}}
	\subfigure[Total Energy]{\includegraphics[width = .4\textwidth]{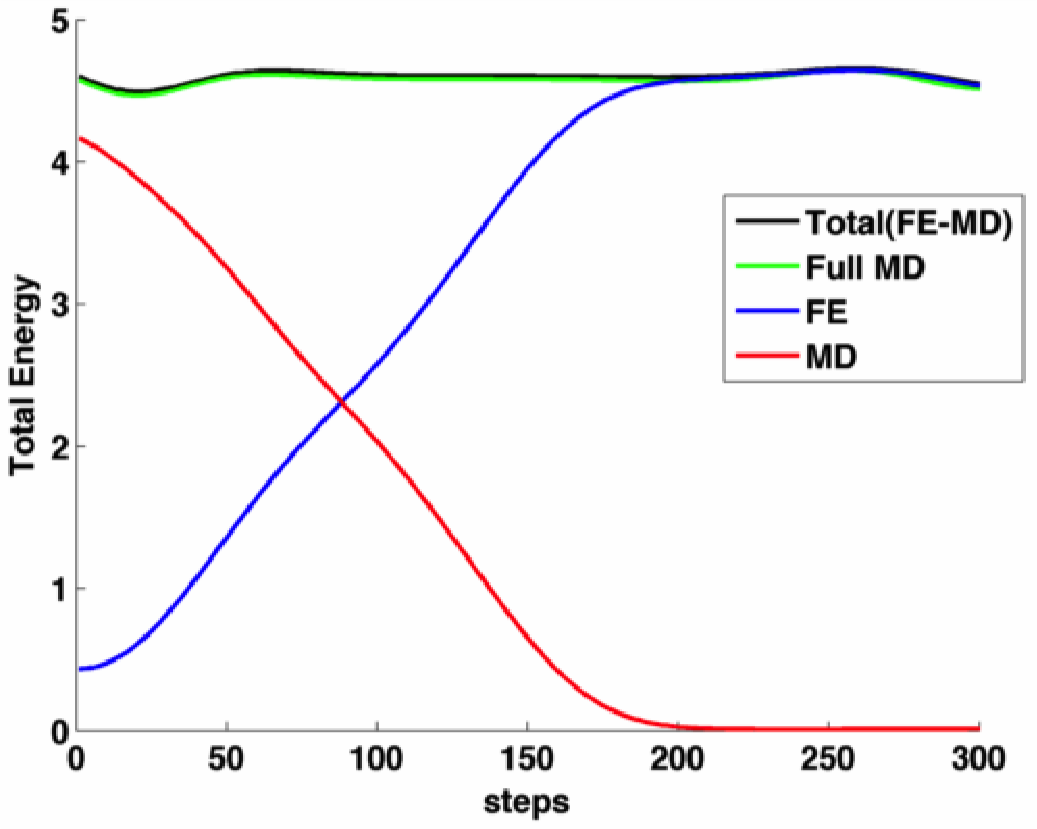}}
	\caption{Energy curves obtained by FE-MD model with Arlequin coupling coefficient obtained by the direct approach. The energy of the entire model (black), which is the summation of the energy of the FE model (blue) and that of the MD model (red), is compared with the energy of a full-MD model (green).}
	\label{fig:eg_energy_alpha}
\end{figure}

\begin{figure}[htp]
	\centering
	\subfigure[Kinetic Energy]{\includegraphics[width = .4\textwidth]{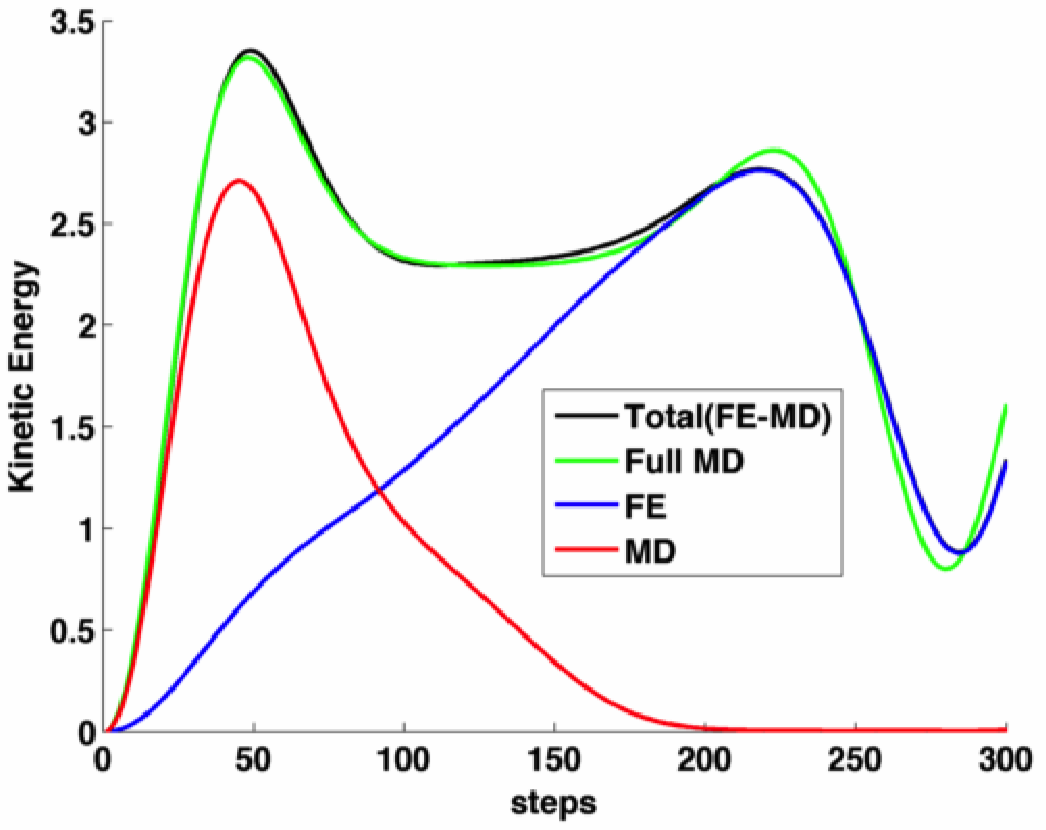}}
	\subfigure[Potential Energy]{\includegraphics[width = .4\textwidth]{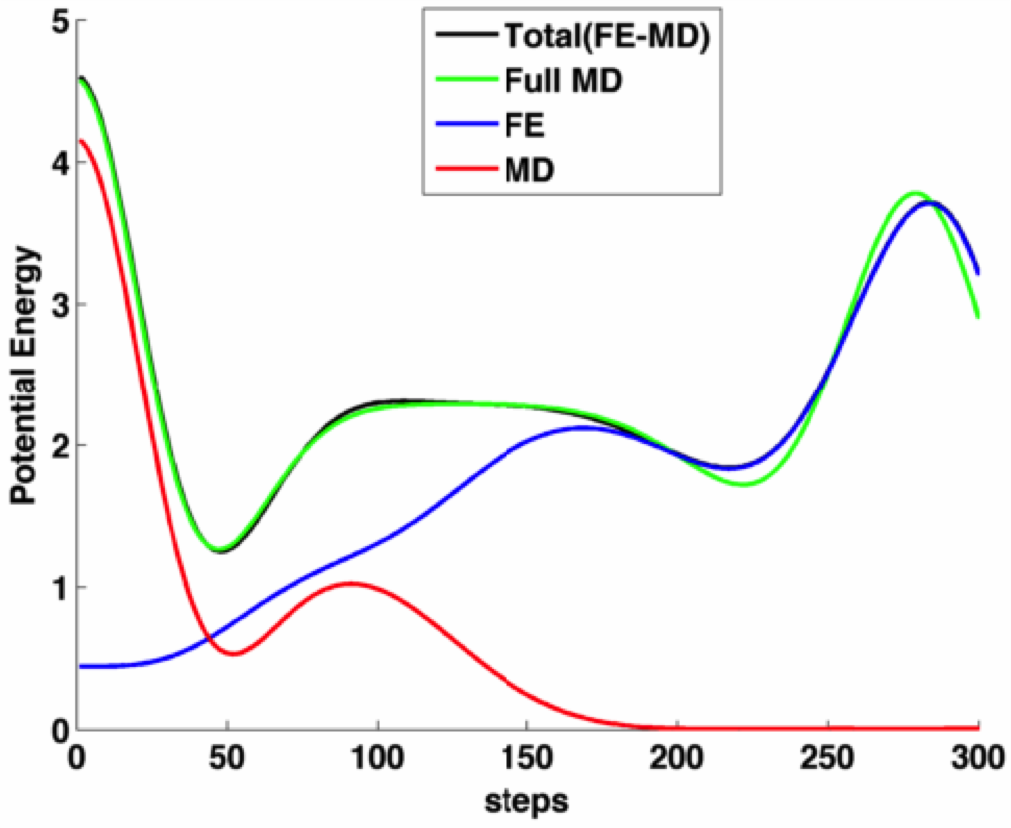}}
	\subfigure[Total Energy]{\includegraphics[width = .4\textwidth]{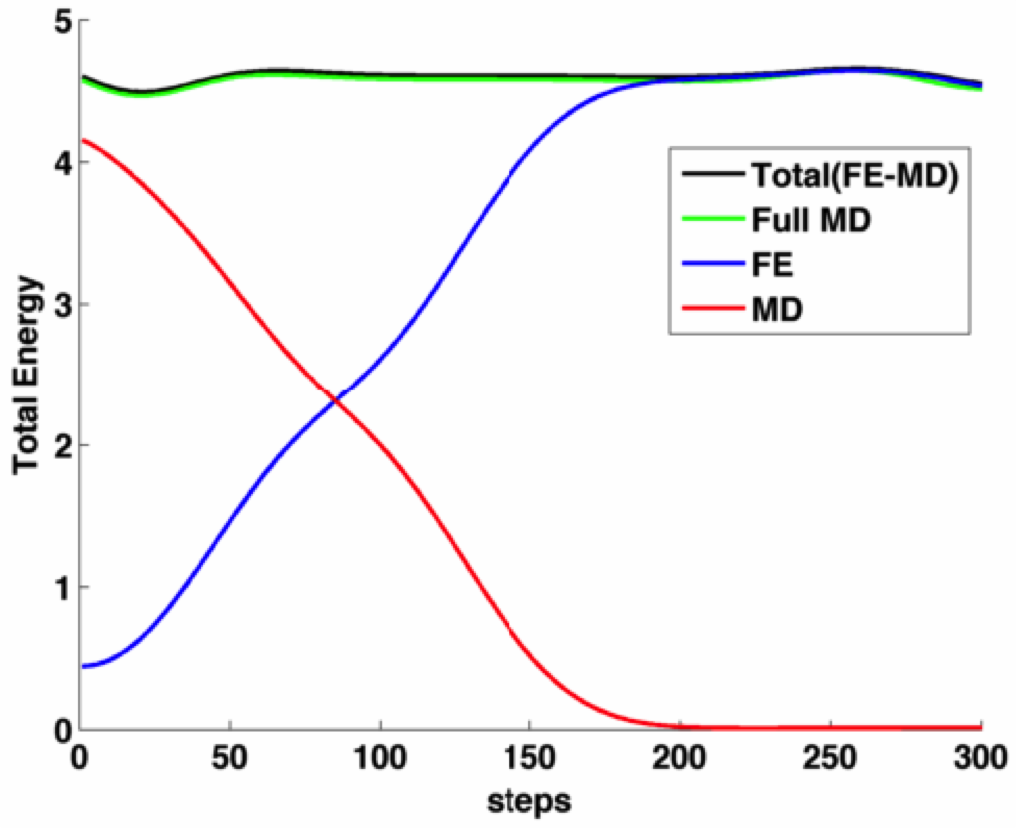}}
	\caption{Energy curves obtained by FE-MD model with Arlequin coupling coefficient obtained by the temperature approach. The energy of the entire model (black), which is the summation of the energy of the FE model (blue) and that of the MD model (red), is compared with the energy of a full-MD model (green).}
	\label{fig:eg_energy_alpha_temp}
\end{figure}

The curves of the kinetic energy, potential energy and total energy for FE-MD models are also plotted in comparison with those for the full-MD model. Fig.~\ref{fig:eg_energy_alpha} and Fig.~\ref{fig:eg_energy_alpha_temp} show the energy curves for the FE-MD model with the coupling coefficient obtained by the direct approach and the temperature approach, accordingly. Both shows very good consistency between the FE-MD model and full-MD model when the coefficient is used, as well as no artificial wave impedance for the transferring of low-frequency motion. Together with the observations from the snapshots discussed previously, we can conclude with good confidence that the Arlequin coupling coefficient computed from the temperature approach works as well as that computed from the direct approach. The advantage of the temperature approach is that it is computationally more efficient because the expensive search algorithm for the boundary points is only performed on the FE nodes on the coupling boundary. Moreover, the result obtained from the temperature approach is completely independent of the geometry of the coupling region, while the result from the direct approach can be slightly different if one chooses the anchor points in a different way. For the temperature approach, we only need the values of the coefficient for the FE nodes on the boundary of the coupling region. No matter how we choose the anchor points, those nodes on the MD side of the coupling boundary would have the value $0$ and the value $1$ on the FE side, while the values inside is determined by the linear thermal conduction law, independent of the choices of the anchor points. \\

\begin{figure}[htp]
	\centering
	\subfigure[Kinetic Energy]{\includegraphics[width = .4\textwidth]{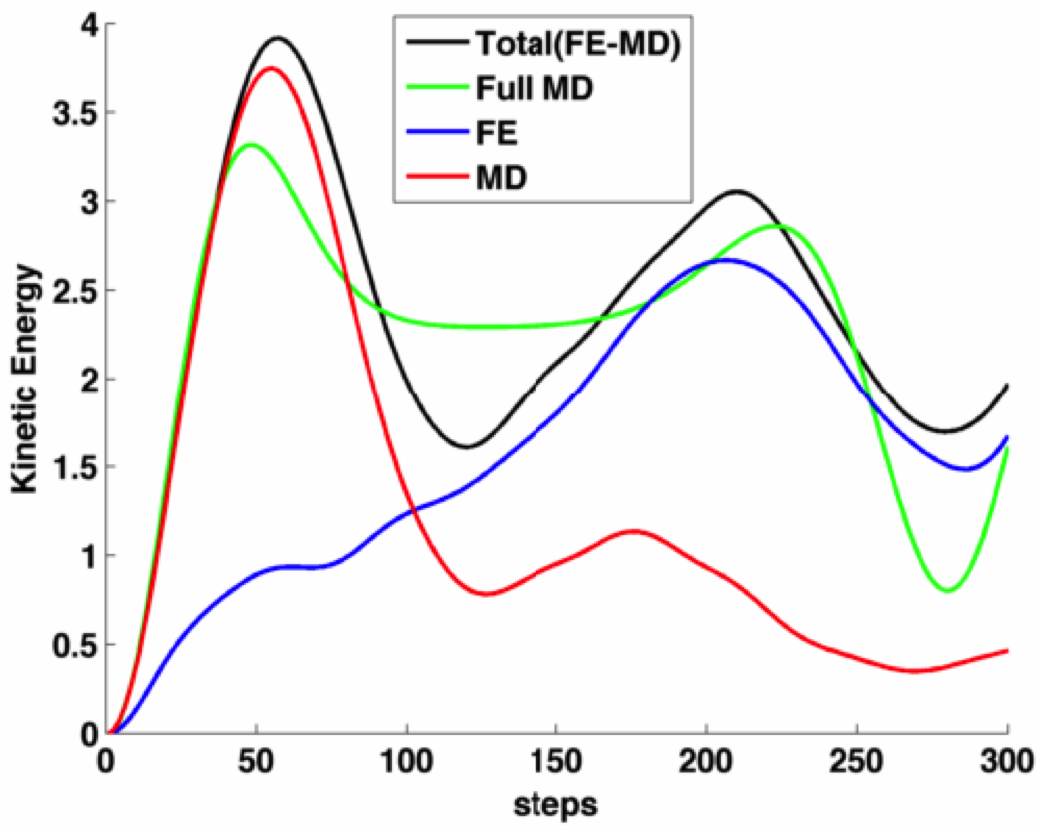}}
	\subfigure[Potential Energy]{\includegraphics[width = .4\textwidth]{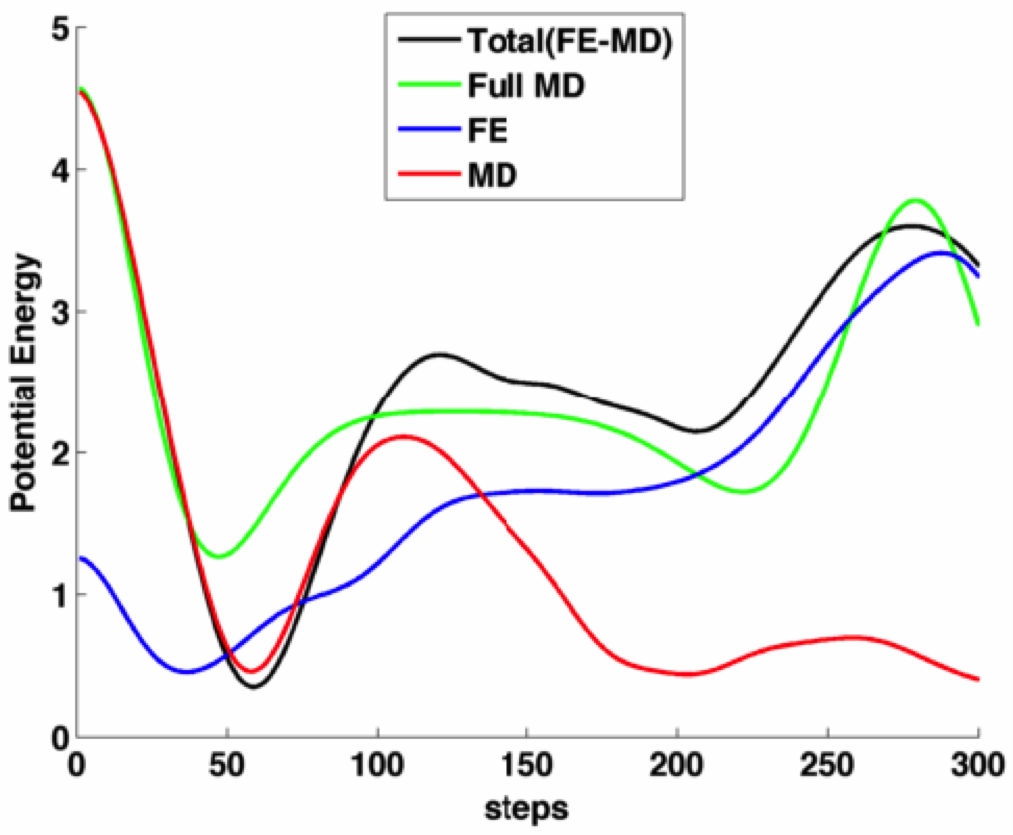}}
	\subfigure[Total Energy]{\includegraphics[width = .4\textwidth]{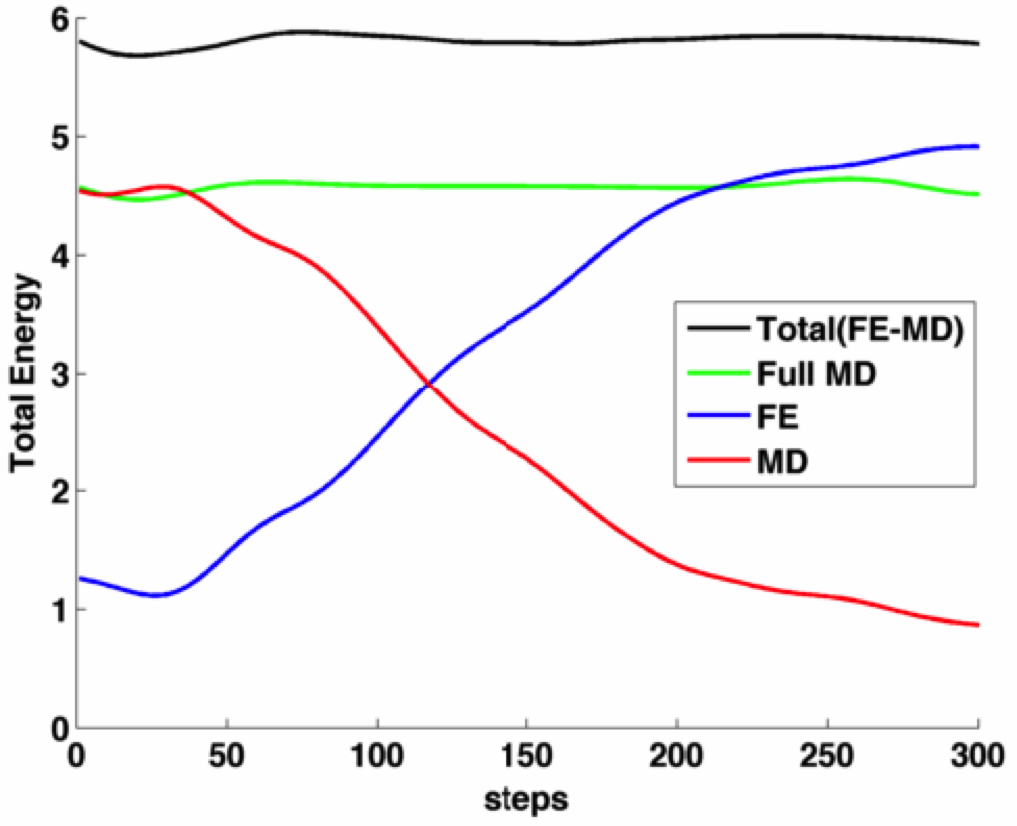}}
	\caption{Energy curves obtained by FE-MD model with no scaling applied to the energies in the coupling region. The energy of the entire model (black), which is the summation of the energy of the FE model (blue) and that of the MD model (red), is compared with the energy of a full-MD model (green).}
	\label{fig:eg_energy_one}
\end{figure}

\begin{figure}[htp]
	\centering
	\subfigure[Kinetic Energy]{\includegraphics[width = .4\textwidth]{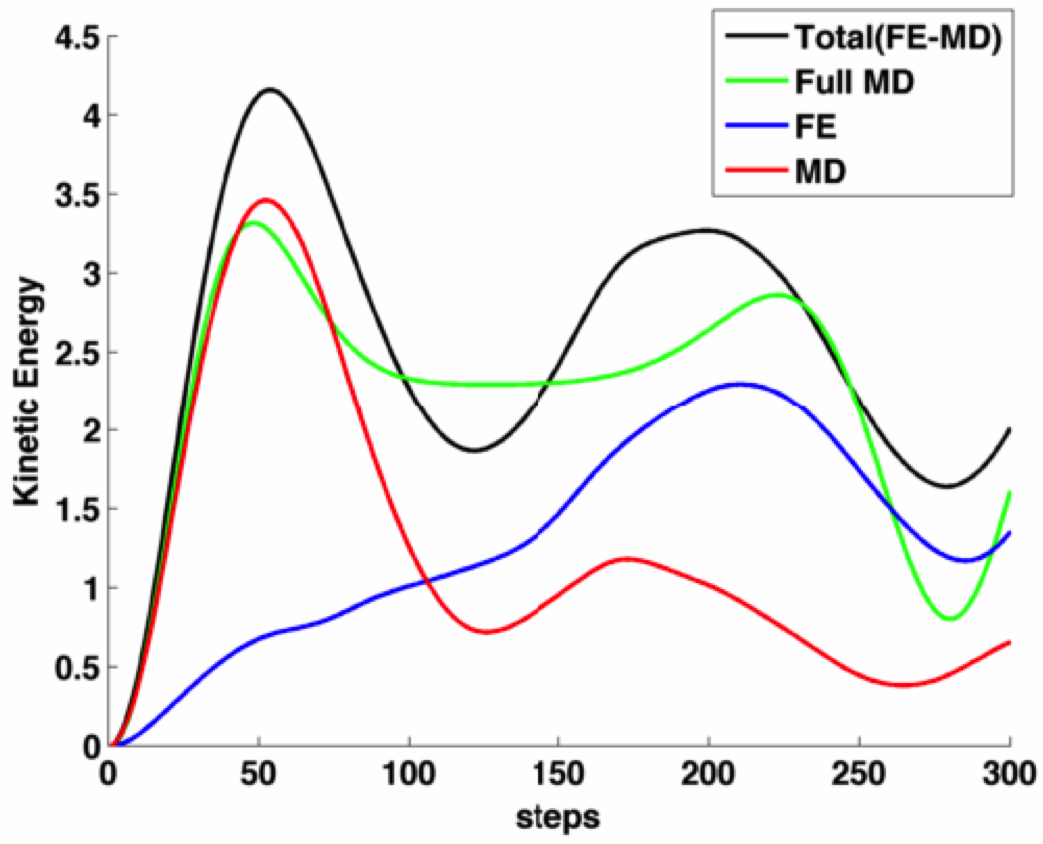}}
	\subfigure[Potential Energy]{\includegraphics[width = .4\textwidth]{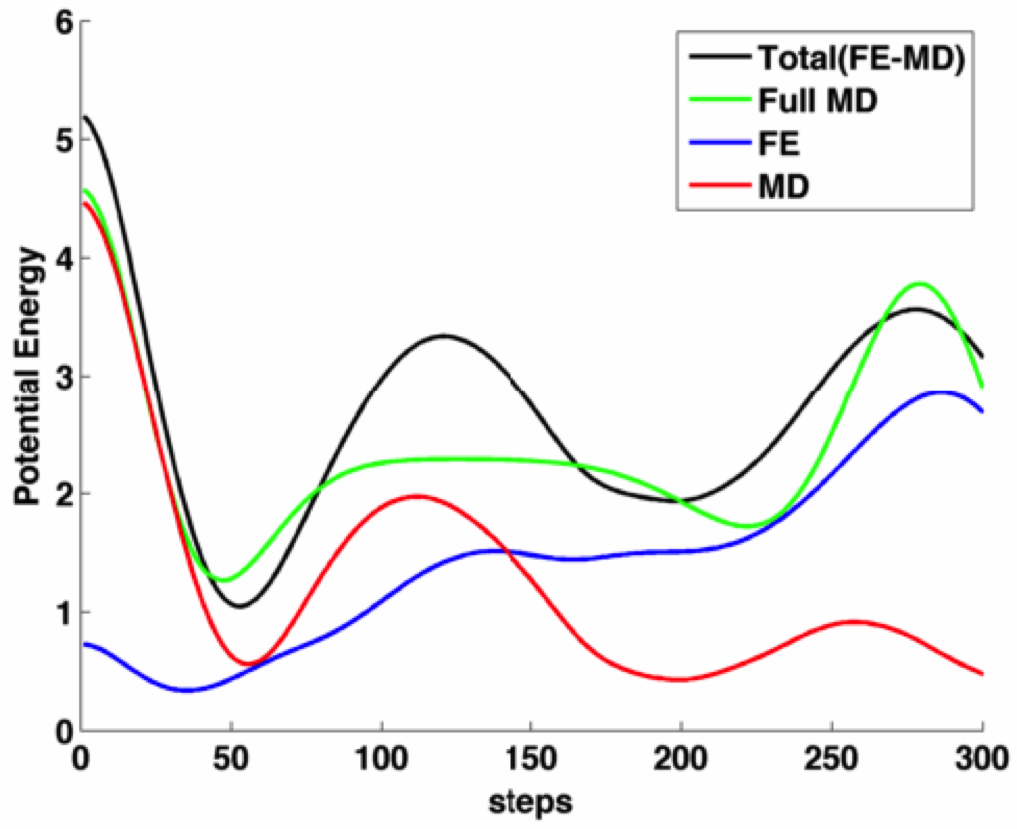}}
	\subfigure[Total Energy]{\includegraphics[width = .4\textwidth]{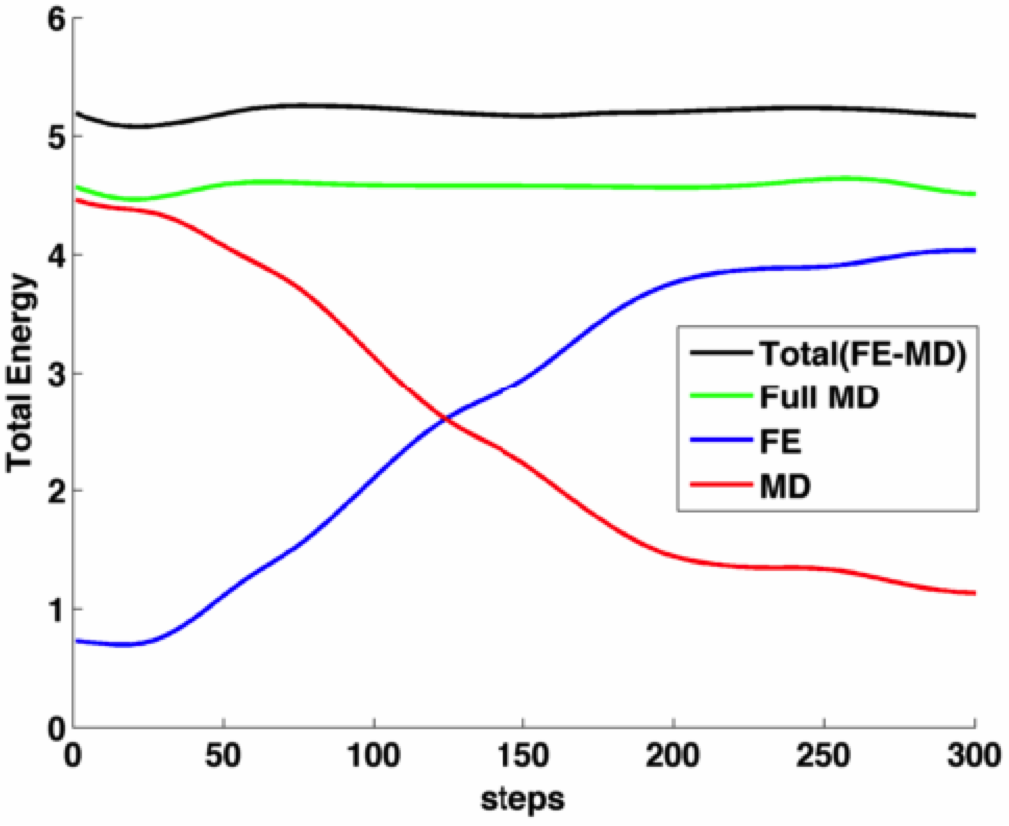}}
	\caption{Energy curves obtained by FE-MD model with constant coupling coefficients: $0.5$ for the MD model and $0.5$ for the FE model. The energy of the entire model (black), which is the summation of the energy of the FE model (blue) and that of the MD model (red), is compared with the energy of a full-MD model (green)}
	\label{fig:eg_energy_half}
\end{figure}

On the other hand, if the energies are not scaled in the coupling region, not only the total energy is over counted in the coupling region, but the overall dynamic behavior of the FE-MD model deviates from the full-MD model significantly, which can be observed in different energy curves plotted in Fig.~\ref{fig:eg_energy_one}. Due to the significant difference in the overall dynamic behavior, the calculation of the total energy cannot be corrected by simply subtracting the duplicated part, either the FE part or the MD part, in the coupling region. Same problems exist, even if the energies in the coupling region are scaled by constant scaling factors Eqn.\eqref{eq:def_alpha_half}, as shown in Fig.~\ref{fig:eg_energy_half}. The constant scaling factor can reduce the amount of double-counted energy in the coupling region by certain amount in comparison to the unscaled case, but the effect is limited. More importantly, constant scaling factors cannot improve the overall dynamic behavior of the system. It can be observed from the energy curves that their deviations from the full-MD curves are at the same level as the unscaled case. Therefore, the scaling factors from the Arlequin coupling coefficient cannot be replaced by constant ones.

\section{Conclusion}
In this work, we introduced a relatively universal procedure to compute the Arlequin coupling coefficient for 2D and 3D problems where the concurrent FE-MD model is preferred. The applicability of the procedure should be independent of the geometry of the coupling region, especially when the temperature approach is chosen. For preprocessing, there are two key steps, which are also the most computationally expensive steps. The first one is the calculation of the iso-parametric coordinates of atoms within corresponding FE elements in the coupling region. The second one is the determination of the boundary points $\mb x_0$ and $\mb x_1$ in Eqn.\eqref{eq:alpha_2D_3D} for calculating the Arlequin coupling coefficient. The direct approach can be quite inefficient because the boundary ray-tracing must be performed on each FE nodes against all surface elements on the coupling boundary. Since we made no assumption on the geometry of the coupling region, it is difficult to localize the searching algorithm. Moreover, the geometry of the coupling region can be very irregular when adaptivity is used, e.g. to trace the propagation of a crack. In such case, additional algorithm is needed to determine proper anchor points inside the MD region. For example, one can choose the points in the central axis of the MD region as anchor points and a point inside the coupling region will be connected to the closest anchor point to construct the ray, or one can divide an irregular region into several sub-regions where one anchor point is sufficient. Another possible drawback on the direct approach would be that the results are dependent on the choice of the anchor point. However, such variation is not significant. On the other hand, the temperature approach is much more efficient and stable, since we only need to use the direct approach to determine the coupling coefficients on the boundary of the coupling region, and the values for the points on the MD side would always be $0$ and those on the $FE$ side would always be one, no matter where the anchor point(s) is. Therefore, the results obtained from the temperature approach are unique. Moreover, if the FE nodes on the coupling boundary can be easily identified for whether they are on the MD side or the FE side, the search for the boundary points (boundary ray-tracing) can be avoided completely. The Arlequin coupling coefficient by the direct approach is linear, same as that for 1D problems. The results by the temperature approach is slightly nonlinear at sharp corners of coupling region. Such nonlinearity will be reduced, if the FE mesh is more refined at those places. However, from the numerical example, we observe very small difference between the behaviors of the FE-MD models using different approaches. What really matters is that the coupling coefficient can gradually decrease from $1$ to $0$ from one side to another. In our numerical example, we found that not double counting the energy is not enough for obtaining consistent behavior to the full-MD model, as the behavior of the FE-MD model with constant scaling factors behaves significantly different from the full-MD model. It is also necessary to smoothly reduce the influence of one model when it is getting further and further away to the side of the other model in the coupling region, and such influence should vanish at the other side. The existence of the Arlequin coupling coefficient means more than energy interpolation, but also represents a gradual transition from one type of model to the other. Such phenomenon might be not limited to FE-MD model, but exists in all types of concurrently coupled models by overlapping domains. Preprocessing and implementation for the concurrent FE-MD model are much more complicated than single models, and can be quite an obstacle for researchers in this field. We hope our work can be of some help for such tasks.
\bibliographystyle{plain}
\bibliography{literatur}   
\newpage
\section*{Appendix: Inverse Iso-Parametric Mappings}
\label{sec:inv_iso_coord}
\begin{figure}[htp]
	\centering
	\includegraphics[width = .4\textwidth]{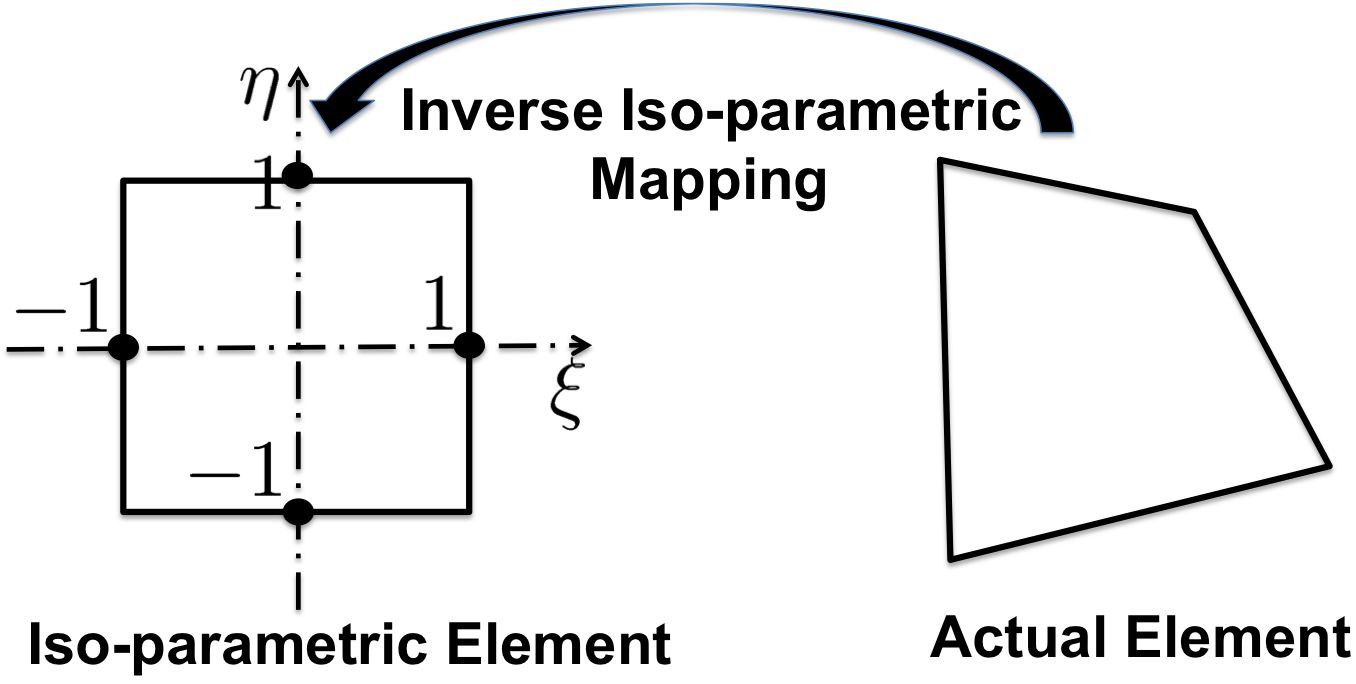}
	\caption{Inverse iso-parametric mapping of a 2D linear quadrilateral element.}
	\label{fig:inv_iso_map}
\end{figure}

To the best of the authors' knowledge, there is no universal approach to perform the inverse iso-parametric mapping. Different approaches must be used for different type of elements. In this subsection, we summarizes the inverse iso-parametric mapping formulas for 5 different types of elements, including 1D bar elements, 2D triangle elements, 2D quadrilateral elements, 3D tetrahedron elements and 3D hexahedron elements. An example of the inverse iso-parametric mapping is demonstrated in Fig.~\ref{fig:inv_iso_map}. \\

\subsubsection{1D Case}
For 1D problems, we only need to determine the iso-parametric coordinates of atoms inside a bar element. Taking an arbitrary element, with nodal coordinates $X_1$ and $X_2$, the iso-parametric coordinate $\xi$ of an atom can be computed by
\begin{equation}
	\xi = \frac{2x - (X_1 + X_2)}{\left|X_2 - X-1\right|}.
	\label{eq:inv_iso_mapping_bar2}
\end{equation}
where $x$ is the Cartesian coordinate of the point. Based on the value of $\xi$, the in/out status of the atom can be determined by
\begin{equation}
	\xi\Rightarrow\left\{\begin{array}{ll}
		-1 \leq \xi \leq 1 & \hbox{in}\\
		\hbox{otherwise} & \hbox{out}
	\end{array}
	\right.
	\label{eq:bar2_in_out}
\end{equation}
\subsubsection{2D Case}
For 2D problems, we consider triangular elements and convex quadrilateral elements. \\

For a \textbf{triangular element}, with the nodal coordinates
\begin{equation}
	\mb X_1 = \left[X_1, Y_1\right],\,\mb X_2 = \left[X_2, Y_2\right]\,,\mb X_3 = \left[X_3, Y_3\right],
\end{equation}
the iso-parametric coordinates $[\xi, \eta, \gamma]$ of an atom with coordinates $[x, y]$ can be determined by
\begin{equation}
	\begin{aligned}
		\xi & = \frac{\left(x-X_3\right)\left(Y_2-Y_3\right) + \left(y-Y_3\right)\left(X_3-X_2\right)}{\left(Y_2-Y_3\right)\left(X_1-X_3\right)-\left(Y_3-Y_1\right)\left(X_3-X_2\right)}\\
		\eta & = \frac{\left(x-X_3\right)\left(Y_3-Y_1\right) + \left(y-Y_3\right)\left(X_1-X_3\right)}{\left(Y_2-Y_3\right)\left(X_1-X_3\right)-\left(Y_3-Y_1\right)\left(X_3-X_2\right)}\\
		\gamma & = 1 - \left(\xi + \eta\right),
	\end{aligned}
	\label{eq:inv_iso_mapping_tri3}
\end{equation}
and the in/out status of the atom can be determined by
\begin{equation}
	\left[\xi, \eta, \gamma\right]\Rightarrow\left\{\begin{array}{ll}
		0 \leq \xi \leq 1\; \&\; 0 \leq \eta \leq 1\; \&\; 0 \leq \gamma \leq 1 & \hbox{in}\\
		\hbox{otherwise} & \hbox{out}
	\end{array}
	\right.
	\label{eq:tri3_in_out}
\end{equation}
where $\&$ denotes the logical \emph{AND} operation. The geometrical positions of nodes $\{1, 2, 3\}$ can be in either clockwise or counter-clockwise direction, while the exact assignment of nodal coordinates does not matter. \\

For a \textbf{quadrilateral element}, the general formula for the inverse iso-parametric mapping is much more involved. Here, we use the formulas derived by Hua \cite{hua1990}. The detailed derivation is quite complicated and interested readers are referred to Hua's original work. Here we summarize the results in a way that can be readily coded. \\
\begin{figure}[htp]
	\centering
	\includegraphics[width = .4\textwidth]{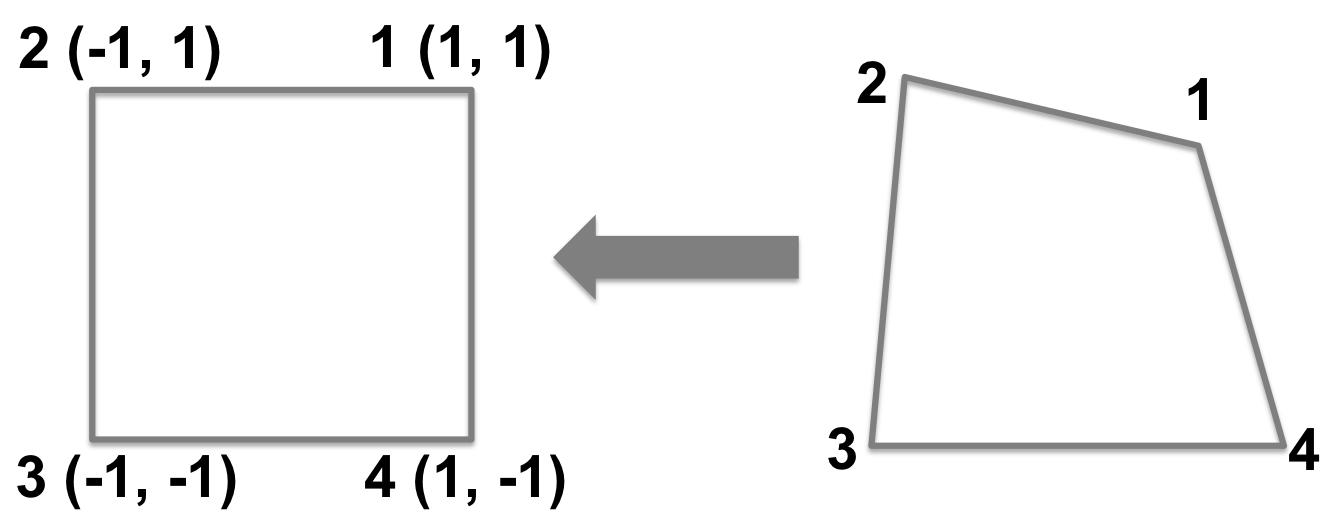}
	\caption{Arrangement of nodes of a quadrilateral element for applying the inverse iso-parametric mapping.}
	\label{fig:inv_iso_quad4_setup}
\end{figure}
Considering an element, with nodal indices arranged in the way shown in Fig.~\ref{fig:inv_iso_quad4_setup} and an atom with coordinates $\left[x, y\right]$ and iso-parametric coordinates $\left[\xi, \eta\right]$, if the atom is inside the element, then the position of the atom can be obtained by
\begin{equation}
	\begin{aligned}
		x & = \sum_{i = 1}^{4}\inv{4}\left(1 + \xi_i\xi\right)\left(1 + \eta_i\eta\right)X_i\\
		y & = \sum_{i = 1}^{4}\inv{4}\left(1 + \xi_i\xi\right)\left(1 + \eta_i\eta\right)Y_i.
	\end{aligned}
	\label{eq:interp_quad4}
\end{equation}
Unlike the triangular case, the iso-parametric coordinates can not be obtained in one step. There are several cases need to be considered. First, we need to define the following constants
\begin{equation}
	\begin{aligned}
		a_1 & = X_1 - X_2  + X_3 - X_4\\
		a_2 & = Y_1 - Y_2  + Y_3 - Y_4\\
		b_1 & = X_1 - X_2  - X_3 + X_4\\
		b_2 & = Y_1 - Y_2  - Y_3 + Y_4\\
		c_1 & = X_1 + X_2  - X_3 - X_4\\
		c_2 & = Y_1 + Y_2  - Y_3 - Y_4
	\end{aligned}
	\label{eq:inv_iso_mapping_quad4_coeff1}
\end{equation}
At certain point of the derivation, one of the above coefficients will be used as denominator,  and therefore we need to consider the cases when they are zero and when they are not zero. In addition, we define the following two constants:
\begin{equation}
	\begin{aligned}
		d_1 & = 4x - \left(X_1 + X_2 + X_3 + X_4\right)\\
		d_2 & = 4y - \left(Y_1 + Y_2 + Y_3 + Y_4\right).
	\end{aligned}
	\label{eq:inv_iso_mapping_quad4_coeff2}
\end{equation}
Now, the solutions for $\left[\xi, \eta\right]$ can be derived as:\\

\noindent\emph{1. When $a_1 = 0$ and $a_2= 0$, then}
\begin{equation}	
	\xi = \frac{d_1c_2 - d_2c_1}{b_1c_1 - b_2c_2}\,\text{and}\,\eta = \frac{b_1d_2 - b_2d_1}{b_1c_1 - b_2c_2}.	
	\label{eq:inv_iso_mapping_quad4_case1}
\end{equation}
\vspace{2pt}\\
\noindent\emph{2. When $a_1 = 0$, $a_2\neq 0$ and $c_1 = 0$, then}
\begin{equation}
	\xi  = \frac{d_1}{b_1}\,\text{and}\,\eta = \frac{b1d_2 - b_2d_1}{a_2d_1 + b_1c_2}.	
	\label{eq:inv_iso_mapping_quad4_case2}
\end{equation}
\vspace{2pt}\\
\noindent\emph{3. When $a_1 = 0$, $a_2\neq 0$ and $c_1 \neq 0$, then}
\begin{equation}
	\begin{aligned}
		&\alpha_1\xi^2 + \alpha_2\xi + \alpha_3 = 0\\
		&\eta = \frac{d_1 - b_1\xi}{c_1},
	\end{aligned}
	\label{eq:inv_iso_mapping_quad4_case3}
\end{equation}
where the coefficients for the quadratic equation are defined as
\begin{equation}
	\begin{aligned}
		\alpha_1 & = a_2b_1\\
		\alpha_2 & = c_2b_1 - a_2d_1 - b_2c_1\\
		\alpha_3 & = d_2c_1 - c_2d_1.
	\end{aligned}
\end{equation}
There will be two solutions for the quadratic equation. Therefore, one need to choose the one with real value and is within the range of $[-1, 1]$. If no solution satisfies the above conditions, it simply means the atom is outside the element. The same criteria applies to all the other cases which involve solving quadratic equations.\\
\vspace{2pt}\\
\noindent\emph{4. When $a_1 \neq 0$, $a_2 = 0$ and $b_2 = 0$, then}
\begin{equation}
	\xi = \frac{d_1c_2 - c_1d_2}{a_1b_2 + b_1c_2}\,\text{and}\,\eta = \frac{d_2}{c_2}.
	\label{eq:inv_iso_mapping_quad4_case4}
\end{equation}
\vspace{2pt}\\
\noindent\emph{5. When $a_1 \neq 0$, $a_2 = 0$ and $b_2 \neq 0$, then}
\begin{equation}
	\begin{aligned}
		&\alpha_1\xi^2 + \alpha_2\xi + \alpha_3 = 0\\
		&\eta = \frac{d_2 - b_2\xi}{c_2}
	\end{aligned}
	\label{eq:inv_iso_mapping_quad4_case5}
\end{equation}
where the coefficients for the quadratic equation are defined as
\begin{equation}
	\begin{aligned}
		\alpha_1 & = a_1b_2\\
		\alpha_2 & = c_1b_2 - a_1d_2 - b_1c_2\\
		\alpha_3 & = d_1c_2 - c_1d_2.
	\end{aligned}
\end{equation}
\vspace{2pt}\\
For the rest of the cases, we define the following additional constants:
\begin{equation}
	\begin{aligned}
		a_b & = a_2b_1 - a_1b_2\\
		a_c & = a_2c_1 - a_1c_2\\
		a_d & = a_2d_1 - a_1d_2.
	\end{aligned}		
\end{equation}
\noindent\emph{6. When $a_1 \neq 0$, $a_2 \neq 0$ and $a_b = 0$, then}
\begin{equation}
	\xi = \frac{d_2a_c - c_1a_d}{b_2a_c + a_2a_d}\,\text{and}\,\eta = \frac{a_d}{a_c}.
	\label{eq:inv_iso_mapping_quad4_case6}
\end{equation}
\vspace{2pt}\\
\noindent\emph{7. When $a_1 \neq 0$, $a_2 \neq 0$, $a_b \neq 0$ and $a_c = 0$, then}
\begin{equation}
	\xi = \frac{a_d}{a_b}\,\text{and}\,\eta = \frac{d_2a_b - b_2a_d}{c_2a_b + a_2a_d}.
	\label{eq:inv_iso_mapping_quad4_case7}
\end{equation}
\vspace{2pt}\\
\noindent\emph{8. When $a_1 \neq 0$, $a_2 \neq 0$, $a_b \neq 0$ and $a_c \neq 0$, then}
\begin{equation}
	\begin{aligned}
		&\alpha_1\xi^2 + \alpha_2\xi + \alpha_3 = 0\\
		&\eta = \frac{a_d - a_b\xi}{a_c}
	\end{aligned}
	\label{eq:inv_iso_mapping_quad4_case8}
\end{equation}
and the coefficients for the quadratic equation are defined as
\begin{equation}
	\begin{aligned}
		\alpha_1 & = a_2a_b\\
		\alpha_2 & = c_2a_b - a_2a_d - b_2a_c\\
		\alpha_3 & = d_2a_c - c_2a_d.
	\end{aligned}
\end{equation}
From the iso-parametric coordinates $\left[\xi, \eta\right]$, the in/out status of the atom with respect to the element can be determined by
\begin{equation}
	\left[\xi, \eta\right]\Rightarrow\left\{\begin{array}{ll}
		0 \leq \xi \leq 1\; \&\; 0 \leq \eta \leq 1& \hbox{in}\\
		\hbox{otherwise} & \hbox{out}.
	\end{array}
	\right.
	\label{eq:quad4_in_out}
\end{equation}
At the end, we need to emphasize again that, to apply the above formulas, the arrangement of the four nodes $\{1, 2, 3, 4\}$ must be the same as what is shown in Fig.~\ref{fig:inv_iso_quad4_setup}, i.e. the correspondence between the coordinates of the 4 corners in the iso-parametric space and those in the reference configuration must match.

\subsubsection{3D Case}
For 3D problems, the tetrahedron elements and hexahedron elements are considered. The ideas behind the derivation are similar to those for 2D cases, but details are more complicated, especially for the hexahedron elements.\\

For a \textbf{tetrahedron element}, with the nodal coordinates
\begin{equation}
	\begin{array}{ll}
		\mb X_1 = \left[X_1, Y_1, Z_1\right], & \mb X_2 = \left[X_2, Y_2, Z_2\right]\\
		\mb X_3 = \left[X_3, Y_3, Z_3\right], & \mb X_4 = \left[X_4, Y_4, Z_4\right],
	\end{array}
\end{equation}
the iso-parametric coordinates $\left[\xi, \eta, \gamma, \zeta\right]$ of an atom with the Cartesian coordinates $[x, y, z]$ can be obtained by
\begin{equation}
	\begin{aligned}
		&\xi = \frac{\det{\mb A_1}}{\det{\mb A}},\,\,\eta = \frac{\det{\mb A_2}}{\det{\mb A}},\,\,\gamma = \frac{\det{\mb A_3}}{\det{\mb A}}\\
		&\zeta = 1 - \left(\xi + \eta + \gamma\right)
	\end{aligned}
	\label{eq:inv_iso_mapping_tet4}
\end{equation}
where $\det{\bullet}$ denotes the determinant operator for matrices $\mb A_i$ and $\mb A$ which are defined as
\begin{equation}
\begin{aligned}
	\mb A_1 = \left[\begin{array}{llll}
		x & y & z & 1\\
		X_2 & Y_2 & Z_2 & 1\\
		X_3 & Y_3 & Z_3 & 1\\
		X_4 & Y_4 & Z_4 & 1
	\end{array}
	\right],\,\,
	\mb A_2 = \left[\begin{array}{llll}
		X_1 & Y_1 & Z_1 & 1\\
		x & y & z & 1\\
		X_3 & Y_3 & Z_3 & 1\\
		X_4 & Y_4 & Z_4 & 1
	\end{array}
	\right]\\
	\mb A_3 = \left[\begin{array}{llll}
		X_1 & Y_1 & Z_1 & 1\\		
		X_2 & Y_2 & Z_2 & 1\\
		x & y & z & 1\\
		X_4 & Y_4 & Z_4 & 1
	\end{array}
	\right],\,\,
	\mb A = \left[\begin{array}{llll}
		X_1 & Y_1 & Z_1 & 1\\
		X_2 & Y_2 & Z_2 & 1\\
		X_3 & Y_3 & Z_3 & 1\\
		X_4 & Y_4 & Z_4 & 1
	\end{array}
	\right].
\end{aligned}	
\end{equation}
The in/out status of the atom is determined by
\begin{equation}
	\left[\xi, \eta, \gamma, \zeta\right]\Rightarrow\left\{\begin{array}{ll}
		0 \leq \left\{\xi, \eta, \gamma, \zeta\right\} \leq 1& \hbox{in}\\
		\hbox{otherwise} & \hbox{out}.
	\end{array}
	\right.
	\label{eq:tet4_in_out}
\end{equation}
\vspace{5pt}\\
\indent The inverse iso-parametric mapping for an atom with respect to a \textbf{hexahedron element} is much more complicated than that for a tetrahedron element. In our work, we use the formula derived by Yuan, et. al. \cite{yuan1994}. As for the quadrilateral element, we only summarize the results here, in a way that can be coded directly in computer. For details about the derivation, one is referred to \cite{yuan1994}.\\
\begin{figure}[htp]
	\centering
	\includegraphics[width = .4\textwidth]{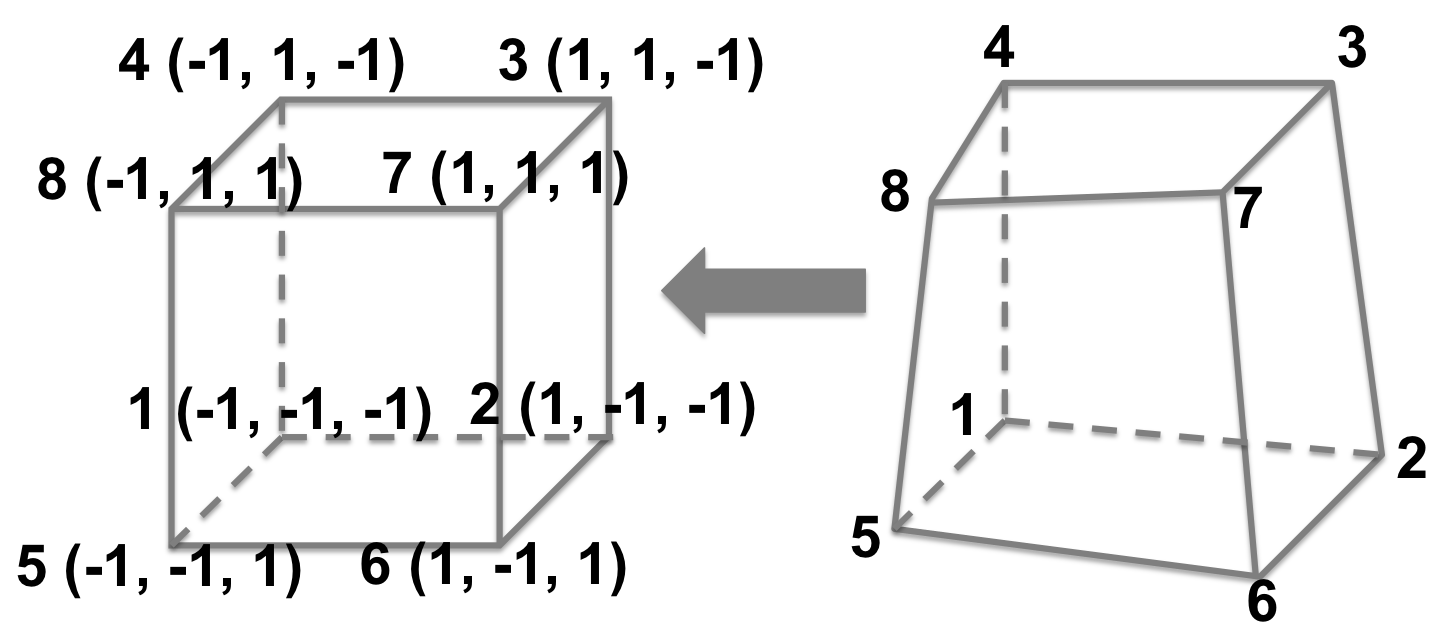}
	\caption{Arrangement of nodes of a hexahedron element for applying the inverse iso-parametric mapping.}
	\label{fig:inv_iso_hex8_setup}
\end{figure}

For the formulas introduced here to be valid, the correspondence between the coordinates of the 8 nodes in the iso-parametric space and the those in the reference configuration must match what is shown in Fig.~\ref{fig:inv_iso_hex8_setup}.  First, we define the following vectors that contain the iso-parametric coordinates of the eight nodes as
\begin{equation}
	\begin{aligned}
		\tens\xi & = \left[\xi_1, \xi_2, \xi_3,  \xi_4,  \xi_5, \xi_6,  \xi_7, \xi_8 \right]\\
		            & = \left[-1, 1, 1, -1, -1, 1, 1, -1\right];\\
		\tens\eta & = \left[\eta_1, \eta_2, \eta_3,  \eta_4,  \eta_5, \eta_6,  \eta_7, \eta_8 \right]\\
		            & = \left[-1, -1, 1, 1, -1, -1, 1, 1\right];\\
		\tens\zeta & = \left[\zeta_1, \zeta_2, \zeta_3,  \zeta_4,  \zeta_5,  \zeta_6,  \zeta_7, \zeta_8 \right]\\
		            & = \left[-1, -1, -1, -1, 1, 1, 1, 1\right],
	\end{aligned}
\end{equation}
as well as the following vectors, $\mb e_z$, $\mb x_z$, $\mb x_e$ and $\mb x_{ez}$, with components
\begin{equation}
\begin{array}{ll}
	\{\mb e_z\}_i = \eta_i\zeta_i,&\{\mb x_z\}_i = \xi_i\zeta_i;\\
	\{\mb x_e\}_i = \xi_i\eta_i,&\{\mb x_{ez}\}_i = \{\mb x_e\}_i\zeta_i.
\end{array}
\end{equation}
Then, we define the following constants
\begin{equation}
	\begin{array}{ll}
		a_1 = \inv{8}\sum_{i = 1}^8 X_i\xi_i, & a_2 = \inv{8}\sum_{i = 1}^8 X_i\eta_i,\\
		a_3 = \inv{8}\sum_{i = 1}^8 X_i\zeta_i, &a_4 = \inv{8}\sum_{i = 1}^8 X_i \{\mb e_z\}_i,\\
		a_5 = \inv{8}\sum_{i = 1}^8 X_i \{\mb x_z\}_i, & a_6 = \inv{8}\sum_{i = 1}^8 X_i \{\mb x_e\}_i,\\
		a_7 = \inv{8}\sum_{i = 1}^8 X_i \{\mb x_{ez}\}_i,		
	\end{array}
\end{equation}
\begin{equation}
	\begin{array}{ll}
		b_1 = \inv{8}\sum_{i = 1}^8 Y_i\xi_i, & b_2 = \inv{8}\sum_{i = 1}^8 Y_i\eta_i,\\
		b_3 = \inv{8}\sum_{i = 1}^8 Y_i\zeta_i, &b_4 = \inv{8}\sum_{i = 1}^8 Y_i \{\mb e_z\}_i,\\
		b_5 = \inv{8}\sum_{i = 1}^8 Y_i \{\mb x_z\}_i, & b_6 = \inv{8}\sum_{i = 1}^8 Y_i \{\mb x_e\}_i,\\
		b_7 = \inv{8}\sum_{i = 1}^8 Y_i \{\mb x_{ez}\}_i	
	\end{array}
\end{equation}
and
\begin{equation}
	\begin{array}{ll}
		c_1 = \inv{8}\sum_{i = 1}^8 Z_i\xi_i, & c_2 = \inv{8}\sum_{i = 1}^8 Z_i\eta_i,\\
		c_3 = \inv{8}\sum_{i = 1}^8 Z_i\zeta_i, &c_4 = \inv{8}\sum_{i = 1}^8 Z_i \{\mb e_z\}_i,\\
		c_5 = \inv{8}\sum_{i = 1}^8 Z_i \{\mb x_z\}_i, & c_6 = \inv{8}\sum_{i = 1}^8 Z_i \{\mb x_e\}_i,\\
		c_7 = \inv{8}\sum_{i = 1}^8 Z_i \{\mb x_{ez}\}_i.		
	\end{array}
\end{equation}
In addition, we need to define the set of vectors $\{\mb e_i\}$ as
\begin{equation}
	\mb e_i = [a_i, b_i, c_i]^T\hspace{10pt}i\,\in\,\{1,2,3,4,5,6,7\}
\end{equation}
the scalar
\begin{equation}
	e_{123} = \mb e_1\cdot\left(\mb e_2\times\mb e_3\right),
\end{equation}
where $\cdot$ denotes the inner product and $\times$ denotes the cross-product; a 3rd order tensor $\mb P$ with components
\begin{equation}
	P_{ijk} = \frac{\mb e_i\cdot\left(\mb e_j\times\mb e_k\right)}{e_{123}},\hspace{5pt}i, j, k\,\in\,\{1,2,3,4,5,6,7\}
\end{equation}
and the matrix
\begin{equation}
	\mb J = \left[\begin{array}{lll}
		a_1 & a_2 & a_3\\
		b_1 & b_2 & b_3\\
		c_1 & c_2 & c_3
	\end{array}
	\right].
\end{equation}
Then, we shift the origin of the atom to the center of the element and the transferred coordinates of the atom $\mb x' = \left[x', y', z'\right]^T$ are
\begin{equation}
	\begin{aligned}
		x' & = x - \inv{8}\sum_{i = 1}^8 X_i\\
		y' & = y - \inv{8}\sum_{i = 1}^8 Y_i\\
		z' & = z - \inv{8}\sum_{i = 1}^8 Z_i.
	\end{aligned}
\end{equation}
In addition, we define the vector
\begin{equation}
	\bar{\tens\xi} = \left[\bar{\xi}_1, \bar{\xi}_2, \bar{\xi}_3\right]^T = \mb J^{-1}\mb x'.
\end{equation}
At the end, another 3 by 3 by 3 third order tensor $\mb G^1$ with nonzero components
\begin{equation}
	\begin{array}{lll}
		G^1_{112} = P_{623}, & G^1_{113} = P_{523},& G^1_{123} = P_{423}\\
		G^1_{212} = P_{163}, & G^1_{213} = P_{153}, & G^1_{223} = P_{143}\\
		G^1_{312} = P_{126}, & G^1_{313} = P_{125}, & G^1_{323} = P_{623}
	\end{array}
\end{equation}
and a 3 by 3 by 3 by 3 fourth order tensor $\mb G^2$ with nonzero components
\begin{equation}
	\begin{aligned}
		G^2_{1123} & = P_{723} - \left[\left(P_{623}P_{143} + P_{523}P_{124}\right)+\left(P_{623}P_{523}\right.\right.\\
		&\left.\left. + P_{423}P_{125}\right) +\left(P_{523}P_{623} + P_{423}P_{163}\right)\right]\\
		G^2_{2123} & = P_{173} - \left[\left(P_{163}P_{143} + P_{153}P_{124}\right)+\left(P_{163}P_{523} \right.\right.\\
		&\left.\left.+ P_{143}P_{125}\right)+\left(P_{153}P_{623} + P_{143}P_{163}\right)\right]\\
		G^2_{3123} & = P_{127} - \left[\left(P_{126}P_{143} + P_{125}P_{124}\right)+\left(P_{126}P_{523} \right.\right.\\
		&\left.\left.+ P_{124}P_{125}\right)+\left(P_{125}P_{623} + P_{124}P_{163}\right)\right],
	\end{aligned}
\end{equation}
\begin{equation}
	\begin{aligned}
		G^2_{1112} & = -2\left(P_{623}P_{163} + P_{523}P_{126}\right)\\
		G^2_{1113} & = -2\left(P_{623}P_{153} + P_{523}P_{125}\right)\\
		G^2_{1221} & = -2\left(P_{623}P_{623} + P_{423}P_{126}\right)\\
		G^2_{1223} & = -2\left(P_{623}P_{423} + P_{423}P_{124}\right)\\
		G^2_{1331} & = -2\left(P_{523}P_{523} + P_{423}P_{153}\right)\\
		G^2_{1332} & = -2\left(P_{523}P_{423} + P_{423}P_{143}\right),
	\end{aligned}
\end{equation}
\begin{equation}
	\begin{aligned}
		G^2_{2112} & = -2\left(P_{163}P_{163} + P_{153}P_{126}\right)\\
		G^2_{2113} & = -2\left(P_{163}P_{153} + P_{153}P_{125}\right)\\
		G^2_{2221} & = -2\left(P_{163}P_{623} + P_{143}P_{126}\right)\\
		G^2_{2223} & = -2\left(P_{163}P_{423} + P_{143}P_{143}\right)\\
		G^2_{2331} & = -2\left(P_{153}P_{523} + P_{143}P_{153}\right)\\
		G^2_{2332} & = -2\left(P_{153}P_{423} + P_{143}P_{143}\right)
	\end{aligned}
\end{equation}
and
\begin{equation}
	\begin{aligned}
		G^2_{3112} & = -2\left(P_{125}P_{126} + P_{126}P_{163}\right)\\
		G^2_{3113} & = -2\left(P_{153}P_{126} + P_{125}P_{125}\right)\\
		G^2_{3221} & = -2\left(P_{124}P_{126} + P_{126}P_{623}\right)\\
		G^2_{3223} & = -2\left(P_{423}P_{126} + P_{124}P_{124}\right)\\
		G^2_{3331} & = -2\left(P_{523}P_{125} + P_{124}P_{153}\right)\\
		G^2_{3332} & = -2\left(P_{423}P_{125} + P_{124}P_{143}\right),
	\end{aligned}
\end{equation}
are defined. With all the variables defined above, the iso-parametric coordinates of the atom can be finally computed as
\begin{equation}
	\begin{aligned}
		\xi & = \bar{\xi}_1 - \inv{2}\sum_{i, j=1}^3 G^1_{1ij}\bar{\xi}_i\bar{\xi}_j - \inv{6}\sum_{i, j, k=1}^3 G^2_{1ijk}\bar{\xi}_i\bar{\xi}_j\bar{\xi}_k\\
		\eta & = \bar{\xi}_2 - \inv{2}\sum_{i, j=1}^3 G^1_{2ij}\bar{\xi}_i\bar{\xi}_j - \inv{6}\sum_{i, j, k=1}^3 G^2_{2ijk}\bar{\xi}_i\bar{\xi}_j\bar{\xi}_k\\
		\zeta & = \bar{\xi}_3 - \inv{2}\sum_{i, j=1}^3 G^1_{3ij}\bar{\xi}_i\bar{\xi}_j - \inv{6}\sum_{i, j, k=1}^3 G^2_{3ijk}\bar{\xi}_i\bar{\xi}_j\bar{\xi}_k
	\end{aligned}
	\label{eq:inv_iso_mapping_hex8}
\end{equation}
with the in/out status determined by
\begin{equation}
	\left[\xi, \eta, \zeta\right]\Rightarrow\left\{\begin{array}{ll}
		-1 \leq \left\{\xi, \eta, \zeta\right\} \leq 1& \hbox{in}\\
		\hbox{otherwise} & \hbox{out}.
	\end{array}
	\right.
	\label{eq:hex8_in_out}
\end{equation}	

\end{document}